\documentclass[acmsmall]{acmart}

\AtBeginDocument{%
  \providecommand\BibTeX{{%
    \normalfont B\kern-0.5em{\scshape i\kern-0.25em b}\kern-0.8em\TeX}}}

\usepackage{tabularx}
\usepackage{graphicx}
\usepackage{xcolor}
\usepackage{multirow}
\usepackage{float}

\usepackage[normalem]{ulem}
\useunder{\uline}{\ul}{}

\newcommand{\rev}[1]{\textcolor{black}{#1}}

\newcommand{\revminor}[1]{\textcolor{black}{#1}}

\newcommand{\del}[1]{}

\newcommand{\delminor}[1]{}

\begin{document}

\setcopyright{acmlicensed}
\acmJournal{PACMHCI}
\acmYear{2023} \acmVolume{7} \acmNumber{CSCW2} \acmArticle{291} \acmMonth{10} \acmPrice{15.00}\acmDOI{10.1145/3610082}

\title[What Actions Do People Want Social Media Platforms to Take on Potentially Misleading Content?]{Remove, Reduce, Inform: What Actions Do People Want \rev{Social Media} Platforms to Take on Potentially Misleading Content?}

\author{Shubham Atreja}
\email{satreja@umich.edu}
\affiliation{%
  \institution{University of Michigan School of Information}
  \country{USA}
}

\author{Libby Hemphill}
\email{libbyh@umich.edu}
\affiliation{%
  \institution{University of Michigan School of Information and ICPSR}
  \country{USA}
}
\author{Paul Resnick}
\email{presnick@umich.edu}
\affiliation{%
  \institution{University of Michigan School of Information}
  \country{USA}
}
\renewcommand{\shortauthors}{Shubham Atreja, Libby Hemphill, \& Paul Resnick}

\begin{abstract}
To reduce the spread of misinformation, social media platforms may take enforcement actions against offending content, such as adding informational warning labels, reducing distribution, or removing content entirely. However, both their actions and their inactions have been controversial and plagued by allegations of partisan bias. 
When it comes to specific content items, surprisingly little is known about what ordinary people want the platforms to do. We provide empirical evidence about a politically balanced panel of lay raters' preferences for three potential platform actions on 368 news articles. 
Our results confirm that on many articles there is a lack of consensus on which actions to take. 
We find a clear hierarchy of perceived severity of actions with a majority of raters wanting informational labels on the most articles and removal on the fewest. 
There was no partisan difference in terms of how many articles deserve platform actions but conservatives did prefer somewhat more action on content from liberal sources, and vice versa.
We also find that judgments about two holistic properties, misleadingness and harm, could serve as an effective proxy to determine what actions would be approved by a majority of raters. 
\end{abstract}




\received{July 2022}
\received[revised]{January 2023}
\received[accepted]{March 2023}

\maketitle

\section{Introduction}

Misinformation and its spread are clearly problems for society. Social media platforms often serve as misinformation conduits and amplifiers \cite{allcott2019trends}. In response, they take a variety of moderation actions against content that they determine to be harmfully misleading. These include, for example, \textit{removing} the content, \textit{reducing} its spread by down-ranking the content, and \textit{informing} readers by attaching a fact-checking result or a warning label to the content \cite{lyonsHardQuestionsWhat2018}. The particular course of action is often decided based on a set of policies (also referred to as guidelines or codebooks) articulated by the platforms, considering content attributes, such as factuality, harm, and speaker intent~\cite{MetaMisinformationStandards,TwitterAddressMisinfo,krishnan2021research}.

However, enforcing actions against harmfully misleading content has proven controversial, at least in the USA \cite{borelFactCheckingWonUs2017, kopitWhyBigTech2021}. People disagree about the harm to individuals or society that may result from widespread exposure to specific content \cite{kopitWhyBigTech2021,bakerChallengesRespondingMisinformation2020}, and a speaker's intention to mislead is not always clear \cite{caplanDeadReckoningNavigating2018}. 
Even trained journals assessing whether news articles are false or misleading have been far from unanimous in their assessments~\cite{allen2020scaling, godel2021moderating, DBLP:journals/corr/abs-2108-07898}. Conservative politicians and pundits have complained that enforcement actions fall disproportionately on publishers expressing conservative viewpoints \cite{borelFactCheckingWonUs2017, lemasterDebateIntensifiesFree}, \rev{while liberal politicians have called for more stringent actions \cite{scott2020despite}.}

Given the controversy, it is surprising how little is known about what ordinary people want the \rev{social media} platforms to do about potentially misleading content. For example, it seems clear that removing content is a more severe action than down-ranking it, but is down-ranking viewed as more severe than adding warning labels, or do people want those two actions to be taken on different content items? Is it even possible for a platform to set a policy that would yield moderation decisions that the vast majority of people would agree with? 
When there is disagreement, is it driven by partisan differences between liberals and conservatives, either in their preferences for moderation in general, or for action against particular articles? Finally, as noted above, \rev{social media} platforms' enforcement policies are based on whether the content is misleading and harmful: to what extent are public's nuanced preferences for moderation actions predictable from those two attributes of the content?
 
We contribute empirical evidence about these questions based on responses to 368 articles by U.S.-based Mechanical Turk raters. We chose articles to over-represent potentially problematic misinformation by selecting popular URLs from problematic sites and URLs that had been flagged by Facebook for further investigation. We collected 54 ratings per article from Mechanical Turk, 18 each from liberals, conservatives, and others. \del{Each article received 18 ratings was rated on Mechanical Turk by 54 raters -- 18 liberals, 18 conservatives, and 18 others. }To ensure that raters were informed about an article, each rater first searched for corroborating information, then judged how misleading and harmful the article was, and finally expressed preferences for what moderation actions, if any, \rev{social media} platforms should take. Prior studies have already assessed and verified the quality of lay raters' \rev{misinformation} judgments by comparing them with expert raters (i.e., journalists) ~\cite{DBLP:journals/corr/abs-2108-07898, allen2020scaling}. 
This paper specifically focuses on understanding preferences for moderation actions, whether there is consensus on those preferences, and whether they can be predicted from judgments of whether articles are misleading and harmful. We find that respondents considered \textit{inform} to be less severe than \textit{reduce}, with \textit{remove} considered most severe. We find a lack of consensus among raters' action preferences on many articles. \rev{For instance, on 146 articles out of 368, the proportion of raters who wanted social media platforms to take \textit{inform} action fell between 30-70\%.} Majorities of liberal raters and conservative raters \rev{in our study} wanted action on about the same number of articles \rev{included in the study}, but liberals wanted more action on articles from conservative sources and vice-versa. Finally we find that a combination of aggregate misleadingness and harm judgments for an article can predict \del{quite well }which actions, if any, the majority of raters preferred.

\section{Background}
\label{sec:background}
\subsection{Actions \rev{Social Media} Platforms Take to Address Misinformation}
\label{sec:actions}
Platforms employ a broad range of enforcement actions in addition to filtering or removing content~\cite{LoModerationInventory}. For misinformation, one approach is to alert users that content may be misleading through a label, color coding, or text. Twitter refers to this as warning~\cite{twitter-warning-def}. Facebook refers to it as an \textit{inform} action~\cite{lyonsHardQuestionsWhat2018}. Another possible action is to downrank a content item so that it appears later in search results or feeds and thus fewer people encounter it. Facebook refers to this as a \textit{reduce} action.

In order to identify which actions to take and when, \rev{social media} platforms have articulated enforcement procedures. These procedures serve two purposes -- i) they articulate guidelines in terms of content attributes that warrant specific enforcement actions (more details in Section \ref{sec:content-attributes}) , and ii) try to reduce the subjectivity of decisions ~\cite{gillespieCustodiansInternetPlatforms2018,zuckerbergBlueprintContentGovernance2018}. For those attributes that depend on human raters to judge, platforms provide a codebook and rater training materials to reduce the subjectivity in their judgment. For example, Google has published the codebook that it asks paid human raters to follow in evaluating the quality of websites as potential search results~\cite{google-search-codebook}. It runs more than 170 pages and defines attributes ranging from expertise and authoritativeness to having a descriptive and helpful title. If there is very high inter-rater agreement about these attributes, they can be thought of as quasi-objective: the judgment will not depend on the opinions or subjectivity of the particular rater who assesses them. Therefore, such enforcement procedures are limited to providing a definitive outcome (in terms of action(s) to take) and fail to account for any uncertainties associated with the outcome \footnote{Analogously, Nesson has argued that juries in the legal context serve to legitimate outcomes by providing a definitive judgment that in principle should have been made by any other jury considering the same evidence; the system does not permit juries to award 70\% of damages in a lawsuit when they think there is a 70\% chance that recompense is deserved or if 70\% of jurors think so~\cite{nesson1984evidence}.}. 

\subsection{Using Content Attributes in Action Decisions}
\label{sec:content-attributes}
As noted above, \rev{social media} platforms often specify their enforcement guidelines in terms of attributes of the content. For enforcement decisions about misinformation, several content attributes have been proposed. The first, of course, is the factuality of claims made in an item (e.g., ``false information that can be verified as such'')~\cite{shuFakeNewsDetection2017}. Platforms have partnered with independent fact-checking organizations to objectively verify the veracity of factual claims present in a content \cite{anannyPartnershipPressLessons2018}. Some versions of this attribute focus on whether content will be \textit{misleading} to readers, rather than directly on factual inaccuracies (e.g., 
``would likely mislead an average person'')~\cite{MetaMisinformationStandards}.  
Researchers \cite{jrDefiningFakeNews2018,jackLexiconLiesTerms2017} and legal scholars \cite{kleinFakeNewsLegal2017} have argued that factual accuracy is an insufficient attribute, as it does not differentiate between satire, unintentional mistakes, and intentionally fabricated content \cite{caplanDeadReckoningNavigating2018}. Consequently, some guidelines also include the potential to cause \textit{harm} and the author's intent as crucial to deciding the course of action against misinformation \cite{vijayaUpdateOurContinuity2020, zuckerbergPreparingElections2018,twitter-warning-def}. 

More broadly, the Credibility Coalition has defined a large set of more specific content attributes to evaluate the credibility of content. These include factors such as, the representativeness of a headline, different types of logical fallacies, reputation of sources, etc. ~\cite{zhangStructuredResponseMisinformation2018a}. However, results show that raters achieved high inter-rater reliability on very few of the Credibility Coalition's attributes. Among content indicators, the Krippendorf alpha (a measure of agreement between independent ratings) for coding whether the title of an article was representative of its content was 0.37; coding for five different types of logical fallacies all had alpha < 0.5. Among context features, only reputation of citations had an alpha level > 0.67 (0.85), a commonly used threshold among communications scholars for human coding \cite{krippendorf2013}. In summary, \rev{social media} platforms' current procedures for misinformation enforcement are faced with the dual challenge of finding the right content attributes that can determine the final action as well as ensuring high rater agreement on these content attributes.   

\subsection{Crowd Judgments on Misinformation}
As noted in the previous subsections, platforms' enforcement procedures mostly rely on content attribute judgments provided by domain experts (i.e., journalists) or trained raters who are paid by the platform. Alternatively, researchers have argued for involving lay people (e.g., platform users) in evaluating potentially misleading content \cite{kimLeveragingCrowdDetect2018,kimLeveragingVolunteerFact2020,pintoFactCheckingCrowdsourcing2019}. Twitter's Birdwatch project invites a community of users to assess whether particular tweets should have warning labels and compose the content of the warnings~\cite{colemanIntroducingBirdwatchCommunitybased2021}.

Several studies that have evaluated the ability of crowds to judge misinformation content. Pennycook and Rand found that Democratic and Republican crowd workers largely agreed with each other when distinguishing mainstream news sources from hyper-partisan and fake news sources \cite{pennycookFightingMisinformationSocial2019a}. Other studies have compared lay raters' annotations with expert raters' annotations on specific articles. Bhuiyan et al.~\cite{bhuiyanInvestigatingDifferencesCrowdsourced2020} asked raters to judge the overall credibility of 50 climate-related articles and found that panels of up to 25 Upwork workers were not as well-correlated as a panel of three journalists were with a panel of three scientists. Three studies assessed lay rater judgments on sets of news articles against the judgments of professional journalists~\cite{allen2020scaling, godel2021moderating, DBLP:journals/corr/abs-2108-07898}. The studies come to somewhat different conclusions; the most positive result, using the same dataset analyzed in this paper, found that the average judgment of sixteen or more MTurk raters was as good as the average judgment of three journalists~\cite{DBLP:journals/corr/abs-2108-07898}. 


\subsection{Public Preferences For Misinformation Moderation Actions}

Relatively little is known about the public's preferences for misinformation moderation. One recent poll of Americans found that the vast majority thought platforms should try to reduce the spread of misinformation and fake news on their platforms, even 67.2\% of strong Republicans \cite{yang_mosleh_zaman_rand_2022}. Support for reducing the spread of a particular form of misinformation, the QAnon Conspiracy, was somewhat lower, among both Democrats and Republicans.
Another study asked respondents about particular problematic social media posts in four topic areas, including Holocaust denial and anti-vaccination misinformation~\cite{kozyreva_herzog_lewandowsky_hertwig_lorenz-spreen_leiser_reifler_2022}. Respondents reported whether the posts should be removed and whether the posting account should be suspended. Again, large majorities were in favor of action, though again more Democrats than Republicans. As noted above these results were only for the remove action and public preferences about other actions (i.e., inform and reduce) are unknown, as is the level of consensus associated with actions on individual articles.

In this study, we seek \del{the public's }informed opinions about misinformation enforcement actions \rev{from the public}. 
We selected raters through stratified sampling to ensure an equal number of self-identified liberals and conservatives, and raters were randomly assigned to rate items, limiting strategic manipulation. Raters were required to pass a quiz to ensure they were minimally-informed about the meaning of the enforcement actions, which is important because many users of social media platforms have only a vague understanding of prioritized feeds~\cite{eslami2015always}. To ensure that they were informed about each content item, raters were required to search for corroborating evidence and provide a URL to an external site to support their ratings.

\section{Research Questions}
\label{sec:rq}

\rev{Social media} platforms' public descriptions of their misinformation mitigation processes often imply a hierarchy of severity of actions: warnings or information labels are the least severe, reducing the distribution through downranking is more severe, and removing an item entirely is the most severe\footnote{https://help.twitter.com/en/rules-and-policies/manipulated-media}. When evaluating the effectiveness of \textit{inform} as a strategy against fighting online misinformation, i.e., whether it affects a reader's perception about the news , studies have concluded that the effect is probably small \cite{claytonRealSolutionsFake2020}, and thus it makes sense to think of reduced distribution as a more severe enforcement action.

It is not obvious, however, that the public perceives downranking to be more severe than applying information labels. In fact, there may not even be a clear ordering, and people may prefer different actions on different kinds of articles; information labels may be preferred for some kind of articles while content downranking may be preferred on other articles. \rev{In order to draw a distinction between the severity of these actions (as determined by social media platforms or otherwise) and how they are perceived by the public, we refer to the latter as ``perceived severity''.} This leads to our first question:

\newcounter{rqnum}
\setcounter{rqnum}{0}

\begin{itemize}
    \item \textbf{RQ\refstepcounter{rqnum}\label{rq-hierarchy}\therqnum: Is there a hierarchy of perceived severity of actions among informed lay raters? }
\end{itemize}

Given the societal controversy about platforms' misinformation moderation practices \cite{borelFactCheckingWonUs2017,kopitWhyBigTech2021}, a natural question is whether the controversy is due to inherent differences among the public about whether and what action(s) should be taken against particular articles. If there are many articles where there is no consensus on the right actions to take, then no moderation process can yield outcomes that please almost everyone almost all the time. This leads to our second question:

\begin{itemize}
    \item \textbf{RQ\refstepcounter{rqnum}\label{rq-agreement}\therqnum: How much agreement is there, among informed lay raters, about the preferred actions to be taken on potentially misleading articles?}
\end{itemize}

Next, we explore the role of political ideology in raters' preferences for misinformation moderation actions. In the U.S., which is the focus of this study, ideological differences are often cast along a single dimension, from liberal to conservative.

Specific to providing judgments about misinformation, prior research shows that doing independent research before rendering judgments increases the correlation between liberal and conservative raters compared to those ideological raters who are less informed. However, some partisan differences still remain \cite{DBLP:journals/corr/abs-2108-07898}. Thus, we might also expect to see some partisan differences in action preferences. These differences may exist for two different reasons that we cover below. 

First, systematic differences in values between liberals and conservatives may produce differences in action preferences. Surveys based on moral foundations theory show that liberals tend to focus only on harm and fairness, while conservatives also focus on loyalty, authority, purity, and liberty \cite{graham2009liberals, haidt2012righteous}.  
If more conservatives harbor libertarian convictions about the importance of free speech even when it is harmful, we might expect conservatives to prefer that platforms take fewer actions overall. Alternatively, it is possible that even when a majority of conservatives prefer some action to be taken, there is a large group of dissenters leading to less agreement among conservative than among liberal raters.

Based on these arguments, we have two research questions:

\begin{itemize}

    \item \textbf{RQ\refstepcounter{rqnum}\label{rq-agreement-ideology}\therqnum: Is there more or less agreement among conservative raters than among liberal raters about their preferred \rev{social media} platform actions?}
    
    \item \textbf{RQ\refstepcounter{rqnum}\label{rq-ideology-amount}\therqnum: Do conservative raters prefer that \rev{social media} platforms act on fewer articles than liberal raters or vice versa?}
    
\end{itemize}

We noted above that systematic differences in values is one potential source of difference in action preferences. Additionally, differences may appear due to strategic reporting when providing these action preferences. Some content, whether accurate or not, may help to sway public opinion in favor of policies or candidates. Therefore, raters may report a preference for actions that increase the distribution of content favoring their political leaning. Even when there is no difference in overall preference for action, we might expect to see differences in which articles raters prefer \rev{social media} platforms to act on. Thus, we ask:

\begin{itemize}
    \item \textbf{RQ\refstepcounter{rqnum}\label{rq-ideology-items}\therqnum: Do conservative raters prefer action on articles from different sources than liberal raters prefer?}
\end{itemize}

Finally, we consider the extent to which action preferences are correlated with judgment of the holistic attributes of misinformation and harm. Explaining specific enforcement actions to lay raters can be challenging as understanding the differences between different actions requires some understanding of ranking algorithms and curated feeds. Pior research suggests that lay raters may not always understand these algorithms \cite{eslami2015always}. 

Alternatively, if raters' preferences for seemingly abstract actions of inform, reduce, and remove can be predicted from the holistic attributes of misleadingness and harm, it would suggest that it is not necessary to directly elicit preferences for specific actions. If, on the other hand, high-level misinformation and harm judgments explain little of the variance in action preferences, it would indicate that action preferences are based on some other factors that are not captured by those two high-level judgments. To understand how best to elicit actionable information from raters, we ask:

\begin{itemize}
    \item \textbf{RQ\refstepcounter{rqnum}\label{rq-proxy}\therqnum: How well can aggregate judgments of whether an article is misleading and/or harmful predict aggregate preferences for the inform, reduce, and remove actions?}
\end{itemize}

\section{Study and Dataset}
\label{sec:study}
Raters on Amazon Mechanical Turk rated a set of news articles. For each article, each rater provided a judgment of how misleading the article was, and how harmful it would be if people were misinformed about the topic. For each article, each rater also reported three binary action preferences, whether they thought platforms should inform users that the article was misleading, reduce the article's distribution, and/or remove it entirely. Details of the dataset and the study procedure follow.


\subsection{Article set} 
A total of 372 articles were selected, taken from two other studies. As described in \cite{allen2020scaling}, a set of 207 articles were selected from a larger set provided by Facebook that was flagged by their internal algorithms as potentially benefiting from fact-checking. The subset was selected based on a criterion that their headline or lede included a factual claim.

The other 165 articles consisted of the most popular article each day from each of five categories \rev{defined by \citet{godel2021moderating}}: liberal mainstream news; conservative mainstream news; liberal low-quality news; conservative low-quality news; and low-quality news sites with no clear political orientation~\cite{godel2021moderating}. Five articles were selected on each of the 33 days between November 13, 2019, and February 6, 2020. Our study was conducted a few months after that. \rev{In their published report, \revminor{\citet{godel2021moderating}} analyzed results for only 99 of the 165 articles, excluding those from liberal and conservative mainstream sites. \revminor{As mentioned in the paper, their analysis only focused on articles from low-quality news sources as ``virtually all articles from mainstream news sources were labeled \emph{true} by professional fact-checkers, and so would be a relatively trivial task for which a crowdsourced approach is not necessary'' \cite{godel2021moderating}.} The authors also added an additional 36 articles in a second wave, \revminor{but we had already run our study by then so these articles were not assessed in our paper.}}

\rev{Four journalists independently evaluated how misleading the articles were on a scale of 1 (not misleading at all) to 7 (extremely misleading). Taking a mean of their judgments, 92 articles received a misleading score above 5, while 181 articles received a misleading score below 3. The complete distribution of their judgments is visualized in Figures \ref{fig:journalists-mit} and \ref{fig:journalists-nyu} in Appendix~\ref{sec:journalist_dist}.}

From our collection of 372 articles, four articles became unreachable during the course of the study and were removed, leaving a total of 368 articles for the analysis. To provide a sense of the articles, Table~\ref{tab:sample-articles} describes four of them where rater action preferences were not uniform.



\begin{table}[H]
  \begin{table}[H]
\centering
\begin{minipage}{\textwidth} 
\resizebox{\textwidth}{!}{
\begin{tabular}{|p{0.2\textwidth}|p{0.4\textwidth}|p{0.2\textwidth}|p{0.15\textwidth}|}
\hline
Title                                                                                       & About the article                                                                                                                                                                                                                                                                                                                                                                                                                                                                                                                                                                                                                                                                       & Action preferences                                                                  & Judgments\footnote{\rev{These judgments were collected on a scale of 1 to 7. The rating process is described in detail in Section \ref{sec:judgments-rating-process}}}                                                            \\ \hline
GOP Removes Sole Polling Place From Famous Hispanic Majority City in Kansas                 
& The headline of the article suggests that the polling station was removed from one of the cities in Kansas. The body of the article however expands to state that the polling station was moved outside of the city center, far from any bus stops.                                                                                                                                                              
& \begin{tabular}[c]{@{}l@{}}Inform: 45\%\\ Reduce: 30\%\\ Remove: 9\%\end{tabular}  & \begin{tabular}[c]{@{}l@{}}Misleading: 2.9\\ Harm: 3.8\end{tabular} \\ \hline
Ginsburg Can't Remember 14th Amendment, Gets Pocket Constitution from the Audience          
& The article refers to an actual incident where Justice Ruth Bader Ginsburg was asked a question about the 14th amendment where she referred to a printed copy of the Constitution before answering. The incident is depicted as it happened. Perhaps, it was politicized a bit by adding the following text ``some of our Supreme Court justices care more about politics and logical gymnastics than the text of the Constitution.’’                         
& \begin{tabular}[c]{@{}l@{}}Inform: 77\%\\ Reduce: 48\%\\ Remove: 16\%\end{tabular} & \begin{tabular}[c]{@{}l@{}}Misleading: 3.5\\ Harm: 3.1\end{tabular} \\ \hline
Bill Clinton: `Allegations Of Sexual Misconduct Should Disqualify A Man From Public Office' 
& The headline of the article attributes a quote to Bill Clinton and the article goes on to state that the comment was made during an interview with MSNBC amid Justice Kavanaugh's confirmation process. The source of the article — The Babylon Bee — is a satire website and carries this disclaimer on every page.                                                                                                                                
& \begin{tabular}[c]{@{}l@{}}Inform: 78\%\\ Reduce: 50\%\\ Remove: 38\%\end{tabular} & \begin{tabular}[c]{@{}l@{}}Misleading: 6.0\\ Harm: 4.0\end{tabular}     \\ \hline

\end{tabular}
}
\caption{Four articles from our database, and the aggregate action preferences and judgments for each.}
\label{tab:sample-articles}
\end{minipage}
\end{table}
    
\end{table}

    


\subsection{Raters}

\rev{Participation was restricted to raters from the U.S.} Before rating their first article, all Mturkers completed a qualification task in which they were asked to rate a sample article and given two attempts to correctly answer a set of multiple choice questions. We quizzed them on their understanding of the instructions, including the  descriptions of what the inform, reduce, and remove actions did. They then completed an online consent form, a four-question multiple choice political knowledge quiz, and a questionnaire about demographics and ideology. MTurkers who did not pass the quiz about the instructions or did not answer at least two questions correctly on the political knowledge quiz were excluded from the study.

At the completion of the qualification task, MTurkers were assigned to one of three groups based on their ideology. We asked about both party affiliation and ideology, \rev{each on a seven-point scale, using two standard questions (VCF0301 and VCF0803) that have been part of the American National Election Studies (ANES) since the 1950s~\cite{anes-documentation}}.
Mturkers who both leaned liberal and leaned toward the Democratic Party were classified as \emph{liberal}; those who both leaned conservative and leaned toward the Republican Party were classified as \emph{conservative}; others were classified as \emph{others}. Others included Mturkers with centrist ideologies as well as Mturkers whose party affiliation did not match their ideology.

\rev{We set up the study so that article HITs were posted at various times over a span of 10 days. We designed the study to collect 54 ratings on each article, 18 each from liberals, conservatives, and others. We stratified the ratings on each article according to rater ideology as we wanted to explore the effect of raters' ideology on their action preferences.} Each rater could rate an article only once. A rater could rate as many articles as they wanted\rev{, as long as other raters from their ideological group had not finished the ratings.} 

\rev{A total of 2185 Mturkers signed up for the study out of whom 622 completed the qualification task and were eligible for the study. Out of these 622 MTurkers, 500 completed at least one rating task. These 500 MTurkers constitute our rater pool. In our pool of raters, more raters were liberals (247) than conservatives (146) or others (107). This is a reflection of the Mturk worker population as a prior survey found that more MTurk workers are liberal than conservative \cite{levay2016demographic}. Since the ratings on each article were stratified according to rater ideology, the median conservative rater rated more articles (17) than the median liberal rater (11). To account for any effects driven by individual raters, we planned to include random effects \cite{brown2021introduction} for raters in our regression analyses. While one rater rated all 368 articles, no one rated them all in one sitting as the articles were released over a span of 10 days.}

\rev{We also collected raters' age, gender, and education level. Raters were 49\% male and 50\% female (others indicated a non-binary gender or preferred not to say). Roughly 60\% of the raters were in the age group 30-49, and 25\% between 18 and 30. Roughly 60\% raters had at least a Bachelor's degree, while 20\% also had a Master's degree. Compared to the US population, our rater pool was younger and also more educated.}

\del{Table~\ref{tab:recruitment-funnel} also provides descriptive statistics about the distribution of number of articles rated per rater. Table~\ref{tab:ratings-per-rater-by-ideology} shows that more raters were liberal than conservative or others. Since each article had 18 ratings from each rater group, conservative and other raters rated more articles per person than did liberal raters, as shown in the second column.}




\subsection{Rating process}
\subsubsection{Step 1: Evidence} Raters were first asked to read a news article by clicking on the URL. In order to solicit an informed judgment on the article, raters were asked to search for corroborating evidence (using a search engine) and provide a link to that evidence in the rating form\delminor{ (see Figure \ref{fig:screen1})}. \rev{The system had automated checks to ensure that – 1) each entry was a valid URL, 2) it was not from the same website as the original article, and 3) it was not a google search link.}. \revminor{We provide screenshots of the rating interface used during the study in Appendix (Figure \ref{fig:screen1}, \ref{fig:screen2}, \ref{fig:screen3}, \ref{fig:screen4})}

\subsubsection{Step 2: Judgments}
\label{sec:judgments-rating-process}

Then, raters were asked to evaluate ``how misleading the article was'' on a \rev{Likert-type question} going from 1=not misleading at all to 7=false or extremely misleading. The question was designed to solicit a holistic judgment about the article rather than focusing on a fixed set of attributes (such as, a factual claim, the accuracy of the headline, etc.). We also avoided using loaded terms such as, ``fake news'' or ``mis/disinformation'' where users may already have preconceived notions about the term \cite{caplanDeadReckoningNavigating2018}. We also provided raters with an option to say that they did not have enough information to make a judgment, although the option was rarely used (<3\% of judgments across all articles); these ratings were excluded from the final analysis. 

A second question \rev{(also Likert-type)} asked raters to evaluate ``how much harm there would be if people were misinformed about this topic'' on a scale (1=no harm at all to 7=extremely harmful). We framed the question counterfactually to discourage any link between misleading judgments and harm judgments (see Figure \ref{fig:screen2} \revminor{in Appendix}). \revminor{In regards to taking action against potentially misleading content, researchers \cite{jrDefiningFakeNews2018,jackLexiconLiesTerms2017} and legal scholars \cite{kleinFakeNewsLegal2017} have argued that simply focusing on the accuracy or misleadingness of content is not sufficient to determine the appropriate course of action as it doesn’t differentiate between consequential misinformation (e.g., potentially impacting health, safety, or participation in democratic processes) and less consequential misinformation (e.g., celebrity gossip). The potential to cause harm has also been included as an additional factor in platforms’ guidelines against misinformation. We designed the question to account for this distinction so that everything that is misleading is not automatically considered harmful and vice versa. The question asks about the harm from people being misinformed on the topic because we wanted people to assess whether the topic was one where misinformation would be consequential, even if the particular article did not contain misinformation.}

\subsubsection{Step 3: Action Preferences} In the next step, we asked each rater to provide their \emph{personal preferences} for action against each news article. First, a rater was asked whether, in their personal opinion, any action was warranted (Figure \ref{fig:screen3} \revminor{in Appendix}). If they answered yes to that question they were asked three binary questions, one for each of three possible actions, inform, reduce, and remove (Figure \ref{fig:screen4} \revminor{in Appendix}). A rater could answer yes to more than one possible action. The instructions included descriptions of what each action means, as shown in the figures. The order of presenting the inform and reduce options was randomized on a per-rater basis, to account for the possibility that the ordering conveyed an implicit ordering of severity of the actions.\footnote{\rev{Given how the actions were framed, we expected ``remove'' to be the most severe of the three actions. And therefore, we were most interested in knowing how the presentation order impacted the preference for the other two actions. To ensure that we have enough data (and power) to reliably conclude any differences between these conditions, we decided to have only two conditions, and keep the position of ``remove'' action fixed.}} 

Following the questions about action preferences, each rater was asked to predict the action preferences of other raters. Answers to those questions are not analyzed in this paper. 

\rev{Raters spent a median of 3 minutes and 50 seconds on rating each article. They were paid \$1 per article, yielding an effective pay rate of just over \$15 per hour, which was our target. } The study was approved by the Institutional Review Board.

\rev{\section{Analysis and Results}}

\label{sec:analysis}

\rev{We organize this section according to our research questions. For each research question, we begin by describing the analysis we conducted followed by the results we found.}

\begin{table}[]
\resizebox{0.9\textwidth}{!}{
\begin{tabular}{p{0.5\textwidth}|p{0.25\textwidth}|l|l}
\textbf{Research Question}                                                                                                                                               & \textbf{Dependent variable}                               & \textbf{Independent variable(s)}                                                                                  &\textbf{Random effects }      \\ \hline
RQ\ref{rq-agreement-ideology}: Is there more or less agreement among conservative raters than among liberal raters about their preferred \rev{social media} platform actions?                                & Does the rater agree with the majority (yes/no)? & rater ideology                                                                                                          & rater id \& article id                  \\ \hline
RQ\ref{rq-ideology-amount}: Do conservative raters prefer that \rev{social media} platforms act on fewer articles than liberal raters?                                                                       & \multirow{2}{*}{rater action preference}         & \multirow{2}{*}{\parbox{0.3\textwidth}{rater ideology \& rater ideology : article source ideology}} & \multirow{2}{*}{rater id \& article id} \\ \cline{1-1}
RQ\ref{rq-ideology-items}: Do conservative raters prefer action on articles from different sources than liberal raters?                                                                  &                                                  &                                                                                                                          &                                            \\ \hline
RQ\ref{rq-proxy}: How well can aggregate judgments of whether an article is misleading and/or harmful predict aggregate preferences for the inform, reduce, and remove actions? & aggregate action preference                      & \parbox{0.3\textwidth}{aggregate misleading judgment + aggregate harm judgment}                                                                  &     --------                                       \\
\end{tabular}
\vspace{5px}
}
\caption{Summary of regression analyses}
\label{tab:reg-analysis}
\end{table}

\subsection{Hierarchy of \rev{Perceived} Severity}

\textbf{RQ\ref{rq-hierarchy}: Is there a hierarchy of perceived severity of actions among informed lay raters?}

\rev{\textbf{Analysis:}} To answer RQ\ref{rq-hierarchy}, we examine the \rev{raters' aggregate action preferences on individual articles.} \del{perceptions revealed by the aggregate preferences for actions on particular articles. }While it would be possible to survey people directly about their general perceptions of the severity of different actions, their responses to an abstract question might not match their revealed preferences in response to specific questions. It could even be that their preferences are different for different articles.

We examine two descriptive statistics. First, for each action type we compute the \rev{percentage} \del{fraction }of ratings where that action was preferred. \rev{Second, we compare the perceived severity of actions on a per-article basis. For each individual article, we determine the action(s) that were preferred by a majority of the raters.} Since a rater could \rev{prefer}\del{ report that} more than one action\del{ was appropriate}, it is possible the majority prefers more than one action on an article. We then identify the set of articles recommended for each action and the subset relations between these article \rev{sets.} \del{recommended for each action.}

\rev{\textbf{Results:}} Table~\ref{tab:overall-descriptives} \rev{shows that 34.09\% of all ratings preferred the inform action, 27.17\% preferred reduce, and 11.85\% preferred remove.}\del{ provides these statistics, aggregated across all articles.} \rev{The group of raters preferred the inform action more often than the reduce action. Remove was the least preferred action}

\begin{figure}[t]

  \centering
  \includegraphics[width=0.3\textwidth]{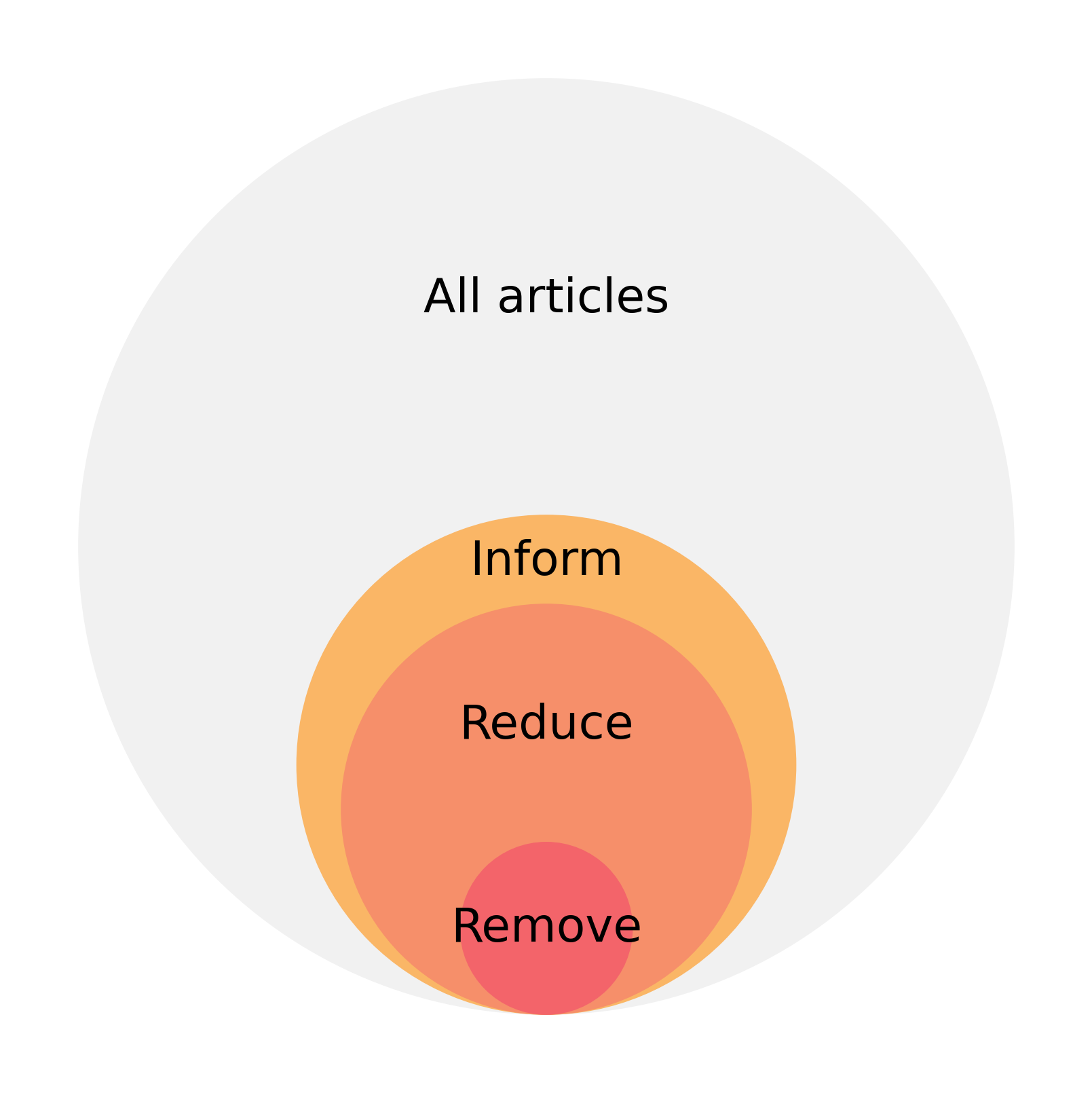}
  \caption{Venn diagram of article sets where majority preferred each action.}%
  \label{fig:rq1}
\end{figure}

Figure \ref{fig:rq1} shows the Venn diagram of the article sets on which a majority of raters preferred each action. For all articles where a majority wanted the reduce action, a majority also wanted the inform action. This suggests a clear hierarchy of \rev{perceived} severity between different actions, where \textit{remove} is perceived as most severe, followed by \textit{reduce}, and then \textit{inform}.

\textbf{\rev{Presentation order of inform and reduce:}} We also compute these same descriptive statistics separately for those raters who were presented with the inform option first and those raters who were presented with the reduce option first to \rev{examine} whether the hierarchy of perceived severity of different actions is robust to their presentation order. 

 
 Columns 2 and 3 of Table ~\ref{tab:presentation-order} \rev{show that the frequency of reduce action increased from 25.35\% to 28.99\% when it was presented before inform.} \del{we see that presenting \textit{reduce} before \textit{inform} increased the frequency of reduce -- 28.99\% compared to 25.35\% when \textit{inform} was presented before \textit{reduce}. }However, inform was still the most frequently preferred action in both conditions. Furthermore, the \rev{subset relations we observed in Figure \ref{fig:rq1} were consistent across the two conditions,} \del{Venn diagram of the article sets remained the same for the two conditions, }i.e, whenever the majority wanted a reduce action, they also wanted an inform action, irrespective of which action was presented first. Therefore, we conclude that the hierarchy of \rev{perceived} severity of different actions remains robust to their presentation order. 

\begin{table}[t]
\begin{tabular}{l|c|c|c}
       & \textbf{All raters} & \textbf{Liberals} & \textbf{Conservatives} \\ \hline
Inform & 34.09      & 37.01     & 35.50       \\ \hline
Reduce & 27.17      & 30.06    & 27.40        \\ \hline
Remove & 11.85      & 13.98   & 11.61    \\   
\end{tabular}
\caption{Breakdown by ideology in fraction of ratings where each action was preferred, aggregated across all articles.}
\label{tab:overall-descriptives}
\end{table}

\begin{table}[t]
\begin{tabular}{c|c|c|c}
       & \textbf{All raters} & \textbf{Inform first} & \textbf{Reduce first} \\ \hline
Inform & 34.09      & 34.42     & 33.80       \\ \hline
Reduce & 27.17      & 25.35    & 28.99        \\ \hline
Remove & 11.85      & 11.10   & 12.60        \\
\end{tabular}
\caption{Breakdown by presentation order in fraction of ratings where each action was preferred, aggregated across all articles.}
\label{tab:presentation-order}
\end{table}

\subsection{(Lack of) Consensus on Preferred Actions}
\textbf{RQ\ref{rq-agreement}: How much agreement is there, among informed lay raters, about the preferred actions to be taken on potentially misleading articles?}

\rev{\textbf{Analysis:}} We compute the aggregate action preferences for each article as the percentage of users who said each action should be taken. We then provide a visual representation of the distribution across all articles. If the distribution of these aggregate preferences is bimodal, with almost every article having close to 0\% or close to 100\% raters wanting each action, then we would have very high agreement among the raters.

In addition to the visual representation, we also report the mean disagreement with the majority-preference for each action.\del{ The disagreement is} We define \rev{disagreement} as the percentage of all raters who do not agree with the majority's preference --- a number in the range $[0,50)$.

\begin{figure}[bt]
  \centering
  \includegraphics[width=1\textwidth]{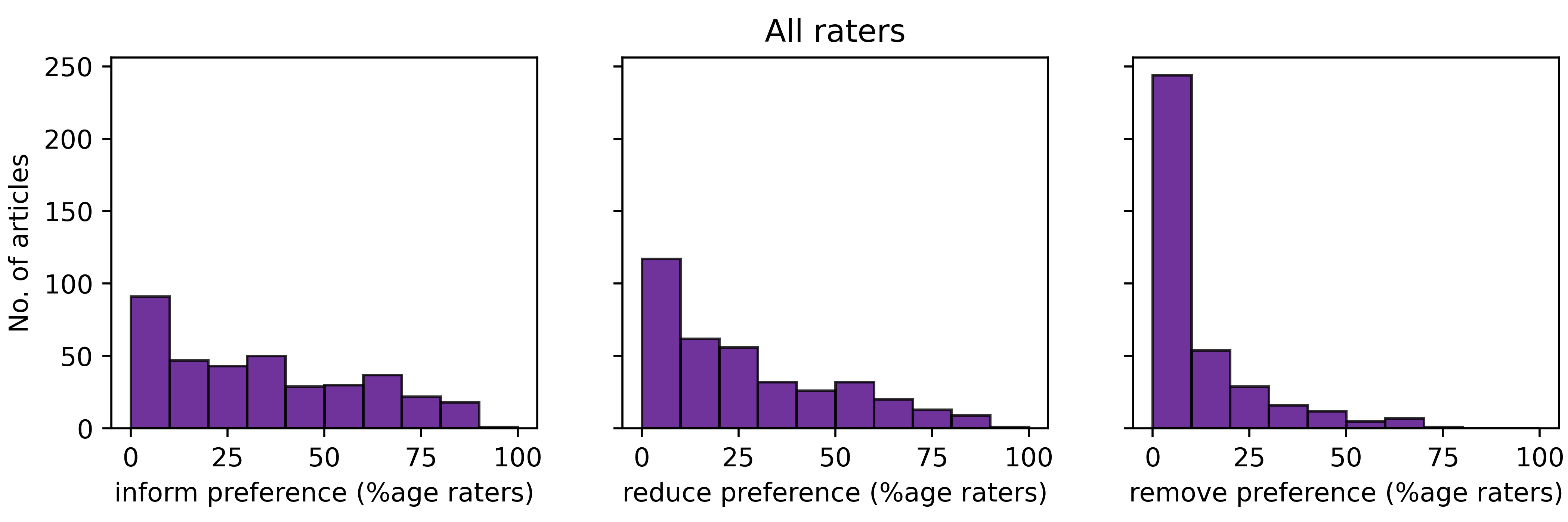}
  \caption{Distribution of raters' aggregate preferences for each action type}%
  \label{fig:dist-of-prefs}
\end{figure}

\rev{\textbf{Results:}} Figure~\ref{fig:dist-of-prefs} shows histograms of aggregate action preferences across all 368 articles. For inform action, 62 articles were present in the middle region of the histogram (i.e., where 40-60\% of the raters wanted the action), indicating high disagreement between raters on these articles. If we expanded the middle range to 30-70\%, the number of articles increased to 146. Similar patterns were observed for reduce action as well. 

When the remove action was preferred by a majority, \del{it was almost always a slim majority, with }only one article achieved a supermajority of more than 70\% raters. Generally, for all three action types, among articles where the majority preferred the action (i.e., the right half of the histograms), it was more common to have 50-70\% of raters prefer the action than to have near universal agreement of 80-100\%. 

Table~\ref{tab:mean-disapproval} shows the mean level of disagreement with the majority preference for each action. Averaging across articles, $23.57\%$ raters disagreed \rev{when the majority preferred} \del{with the majority preference about whether }to take the inform action. \rev{For remove action,} the mean level of disagreement was \rev{11.05\%} \del{lower for the remove action }because there were many articles that almost everyone agreed should not be removed (see histogram for remove action in Figure \ref{fig:dist-of-prefs}). Note that the mean disagreement would only be marginally higher (11.85\% instead of 11.05\%) with a default decision of not removing any article.

\begin{table}[t]
\begin{tabular}{l|c|c|c}
            & \textbf{Inform} & \textbf{Reduce} & \textbf{Remove}        \\ \hline
Disagreement & 23.57 & 21.12   & 11.05        \\
\end{tabular}
\caption{Mean across articles, of percentage of raters who disagreed with the majority preference.}
\label{tab:mean-disapproval}
\end{table}

We considered computing an inter-rater reliability measure such as Krippendorf's alpha~\cite{krippendorf2013}. However, it is not clear how one would interpret the result. As noted in the background section, heuristic thresholds (e.g., 0.67) on these reliability measures are sometimes used to separate attributes with high agreement from low agreement attributes. However, to assess the extent of consensus on preferred actions, visualizing the distribution of their aggregate preferences across articles and computing the mean as a summary statistic provides more intuition. 

\subsection{Partisan Differences in Preferences}

\textbf{RQ\ref{rq-agreement-ideology}: Is there more or less agreement among conservative raters than among liberal raters about their preferred \rev{social media} platform actions?}

\rev{\textbf{Analysis:}} We conduct regression analysis to identify whether liberal and conservative raters have different rates of agreement with other raters of their own ideology. We train a separate model for each action type. The dependent variable in a model represents whether an individual rater's action preference for a given article matches the majority-preference of raters of their own ideology. Since the dependent variable is binary, we use Logistic Regression. The main independent variable is the rater's ideology. 
The dataset consists of the 13004 ratings from the raters who were either classified as conservatives or liberals, excluding the other raters. 
Since we have ratings from multiple raters on each article, and most raters have rated multiple articles, we include random effects (constant slope) \rev{\cite{brown2021introduction}} for rater id and article id in our model to account for any rater-specific or article-specific bias.


\rev{\textbf{Results:}} Table \ref{tab:reg-agreement} shows the regression coefficients from our model. Except for the inform action, there were no significant differences between conservatives and liberals in their level of agreement with the majority-preference of their respective groups. For inform, conservatives were significantly less likely \rev{(p<0.01)} to agree with the majority-preference of their group, compared to liberals.

\begin{table}[!htbp] \centering 
 
\begin{tabular}{@{\extracolsep{5pt}}lccc} 
\\[-1.8ex]\hline 
\hline \\[-1.8ex] 
 & \multicolumn{3}{c}{\textit{Dependent variable:}} \\ 
\cline{2-4} 
 & Inform preference & Reduce preference & Remove preference \\ 
\\[-1.8ex] & (1) & (2) & (3)\\ 
\hline \\[-1.8ex] 
 conservative & $-$0.267$^{***}$ & $-$0.149 & 0.072 \\ 
  & (0.085) & (0.088) & (0.153) \\ 
  & & & \\ 
 Intercept & 1.685$^{***}$ & 1.816$^{***}$ & 3.133$^{***}$ \\ 
  & (0.071) & (0.078) & (0.133) \\ 
  & & & \\ 
\hline \\[-1.8ex] 
Random effects &  &  &  \\ 
sd(rater\_id) & 0.493 & 0.508 & 0.952 \\ 
sd(article\_id) & 0.800 & 0.968 & 1.487 \\ 
Observations & 13,004 & 13,004 & 13,004 \\ 
Log Likelihood & $-$6,374.042 & $-$5,945.140 & $-$3,548.398 \\ 
\hline 
\hline \\[-1.8ex] 
\textit{Note:}  & \multicolumn{3}{r}{$^{**}$p$<$0.05; $^{***}$p$<$0.01} \\ 
\end{tabular} 

 \caption{Regression models for estimating the impact of rater ideology on agreement with the majority preference of their ideology (RQ\ref{rq-agreement-ideology})} 
  \label{tab:reg-agreement} 
  
\end{table}

\begin{table}[!htbp] \centering 
\resizebox{\textwidth}{!}{

\begin{tabular}{@{\extracolsep{5pt}}lccc} 
\\[-1.8ex]\hline 
\hline \\[-1.8ex] 
 & \multicolumn{3}{c}{\textit{Dependent variable:}} \\ 
\cline{2-4} \\
 & Inform preference & Reduce preference & Remove preference \\ 
\\[-1.8ex] & (1) & (2) & (3)\\ 
\hline \\[-1.8ex] 
 conservative & 0.293 & 0.161 & $-$0.069 \\ 
  & (0.169) & (0.175) & (0.248) \\ 
  & & & \\ 
 source\_pro\_conservative & 1.238$^{***}$ & 1.265$^{***}$ & 1.040$^{***}$ \\ 
  & (0.248) & (0.247) & (0.291) \\ 
  & & & \\ 
 source\_pro\_liberal & 0.113 & 0.188 & $-$0.407 \\ 
  & (0.319) & (0.318) & (0.391) \\ 
  & & & \\ 
 source\_unknown & 0.828$^{**}$ & 0.847$^{**}$ & 0.824$^{**}$ \\ 
  & (0.335) & (0.332) & (0.384) \\ 
  & & & \\ 
 conservative : source\_pro\_conservative & $-$1.052$^{***}$ & $-$1.000$^{***}$ & $-$0.760$^{***}$ \\ 
  & (0.118) & (0.124) & (0.175) \\ 
  & & & \\ 
 conservative : source\_pro\_liberal & 0.763$^{***}$ & 0.658$^{***}$ & 0.907$^{***}$ \\ 
  & (0.146) & (0.156) & (0.249) \\ 
  & & & \\ 
 conservative : source\_unknown & $-$0.540$^{***}$ & $-$0.563$^{***}$ & $-$0.215 \\ 
  & (0.160) & (0.164) & (0.219) \\ 
  & & & \\ 
 Intercept & $-$1.769$^{***}$ & $-$2.258$^{***}$ & $-$4.115$^{***}$ \\ 
  & (0.197) & (0.198) & (0.259) \\ 
  & & & \\ 
\hline \\[-1.8ex] 
Random effects &  &  &  \\ 
sd(Rater\_id) & 1.11 & 1.12 & 1.54 \\ 
sd(article\_id) & 1.88 & 1.84 & 1.97 \\ 
Observations & 13,004 & 13,004 & 13,004 \\ 
Log Likelihood & $-$5,936.760 & $-$5,501.412 & $-$3,244.302 \\ 
\hline 
\hline \\[-1.8ex] 
\textit{Note:}  & \multicolumn{3}{r}{$^{**}$p$<$0.05; $^{***}$p$<$0.01} \\ 
\end{tabular}
}
 \caption{Regression Models for estimating the impact of rater ideology and the ideological leaning of the source on rater action preference (RQ\ref{rq-ideology-amount} and RQ\ref{rq-ideology-items}). In all models, `liberal' and `no known bias' are the reference categories.} 
  \label{tab:reg-source-pref} 
  
\end{table} 

\begin{table}[]
\begin{tabular}{l|c|c|c|c}
                  & \textbf{Estimate} & \textbf{SE}    & \textbf{z.ratio} & \textbf{p-value} \\ \hline
Inform  & -0.122   & 0.178 & -0.684  & 0.4939  \\ \hline
Reduce  & -0.291   & 0.184 & -1.585  & 0.1130  \\ \hline
Remove  & -0.103   & 0.269 & -0.384  & 0.7009  \\ 
\end{tabular}

\caption{Reporting the contrast on rater ideology (between conservatives and liberals) averaged across all source ideology levels}
\label{tab:contrasts}
\end{table}

\begin{table}[]
    \begin{tabular}{l|c|c|c}
       & \textbf{All raters} & \textbf{Liberals} & \textbf{Conservatives} \\ \hline
    Inform & 104        & 118           & 108                \\ \hline
    Reduce & 70         & 80            & 74                 \\ \hline
    Remove & 12        & 15            & 22                 \\ 
    \end{tabular}
  \caption{No. of articles recommended for each action type based on the aggregate preferences of different user groups}%
  \label{tab:rq1}
\end{table}

\textbf{RQ\ref{rq-ideology-amount}: Do conservative raters prefer that \rev{social media} platforms act on fewer articles than liberal raters or vice versa?}

\rev{\textbf{Analysis:}} For RQ\ref{rq-ideology-amount} we conducted another regression analysis. We again train separate logistic regression models for each action\del{ type}. The dependent variable is the individual rater's action preference (yes or no). The independent variables include rater ideology, article source ideology, and their interaction terms (details about the article source ideology and its interaction term are revealed in the analysis of RQ\ref{rq-ideology-items} below). Similar to the previous analysis, we also included random effects (constant slope)\rev{\cite{brown2021introduction}} for rater id and article id to account for any rater-specific or article-specific bias. We then conduct a post-hoc marginal means analysis\footnote{https://cran.r-project.org/web/packages/emmeans/vignettes/interactions.html} to identify whether rater ideology has a significant impact on their action preference, averaged across all source ideology labels. 

\rev{\textbf{Results:}} Table \ref{tab:contrasts} shows \rev{the results from the marginal means analysis. We} \del{that, following a marginal means analysis to get an average effect across all source ideology labels, there were }found no significant differences between conservatives and liberals in terms of how many articles they preferred the action, and that holds for all actions \rev{(p>0.1 for all actions).} 

Table~\ref{tab:rq1} further summarizes the majority preferences of liberals and conservatives. While there were minor differences in the number of articles that majorities of liberals and conservatives preferred \rev{social media} platforms act on, the regression analysis reported above showed that none of these differences are statistically significant.

\textbf{RQ\ref{rq-ideology-items}: Do conservative raters prefer action on articles from \emph{different sources} than liberal raters?}

\rev{\textbf{Analysis:}} For RQ\ref{rq-ideology-items}, we use the same regression analysis as above. This time, we observe how the \textit{interaction} between rater ideology and article source ideology impacts the rater's action preference. For article source ideology, we extract labels from MBFC\footnote{https://mediabiasfactcheck.com} which classifies news sources according to their political leaning (pro-liberal, pro-conservative, no bias, or unknown), and apply the site's label to the article. 

\rev{\textbf{Results:}} The coefficients of the interaction terms in Table \ref{tab:reg-source-pref} show the difference from a reference of a liberal rater and a news source labeled as having no known biases. We find that compared to liberal raters, conservatives were significantly less likely to prefer action on articles from pro-conservative sources (see conservative :source\_pro\_conservative in Table \ref{tab:reg-source-pref}\rev{; p<0.01}). Conversely, on articles from pro-liberal sources, conservative raters were significantly more likely than liberal raters to prefer an action \rev{(p<0.01)}. The results hold for all actions. Conservatives were also less likely to prefer inform and reduce actions on articles from sources whose ideological leaning is not available on MBFC\footnote{https://mediabiasfactcheck.com} \rev{(p<0.01)}.

To quantify the differences in a more easily interpretable way, we also compute the correlation, across articles, between the aggregate preferences of conservative raters and liberal raters, computed separately for articles from different sources.\del{ (Table \ref{tab:rq4}).} \rev{Table \ref{tab:rq4} shows that on unbiased sources, the correlation between aggregate preferences of liberals and conservatives falls within 0.91-0.92 for all actions. The correlation drops between 0.64-0.79 on articles from pro-conservative sources. The minimum correlation is 0.55 (remove action on articles from pro-liberal sources)}

\del{While conservative and liberal action preferences were largely correlated on unbiased sources, the correlation dropped quite a bit for sources that are ideologically biased toward either side. We note, however, that positive correlations in the range of 0.55 to 0.80 are still fairly high. }

\begin{table}[]
\begin{tabular}{l|c|c|c}

       & \textbf{Unbiased source} & \textbf{Pro-liberal source} & \textbf{Pro-conservative source} \\ \hline
Inform & 0.92            & 0.66               & 0.79                    \\ \hline
Reduce & 0.91            & 0.68               & 0.80                    \\ \hline
Remove & 0.91            & 0.55               & 0.64                    \\ 
\end{tabular}
\caption{Correlations between percentages of conservatives and liberals who preferred an action, on articles from different types of sources.}
\label{tab:rq4}
\end{table}

\subsection{Reducibility to Misleading and Harm Judgments}
\label{sec:reducibility}

\textbf{RQ\ref{rq-proxy}: How well can aggregate judgments of whether an article is misleading and/or harmful predict aggregate preferences for the inform, reduce, and remove actions?}

\rev{\textbf{Analysis:}} We use another set of regression analyses to show whether the raters' aggregate action preferences can be predicted from their aggregate judgments. We again train a separate model for each action type. The data consists of 368 rows (one per article). The dependent variable for each model is the fraction of raters who wanted that action. Since the dependent variable is a count proportion, we use a Generalized Linear Regression Model (GLM) with binomial family and a logit link (as recommended by Zuur et al. \cite{zuurGLMGAMAbsence2009}).

The independent variables are the means of raters' misleading and harm judgments. 
We fit four models using different combinations of independent variables -- only misleading judgments, only harm judgments, both without an interaction term, and both plus an interaction term. We use information criteria (AIC and BIC) to select the best fit model for each action type. 

\begin{figure}
    \centering
    \includegraphics[width=0.7\textwidth]{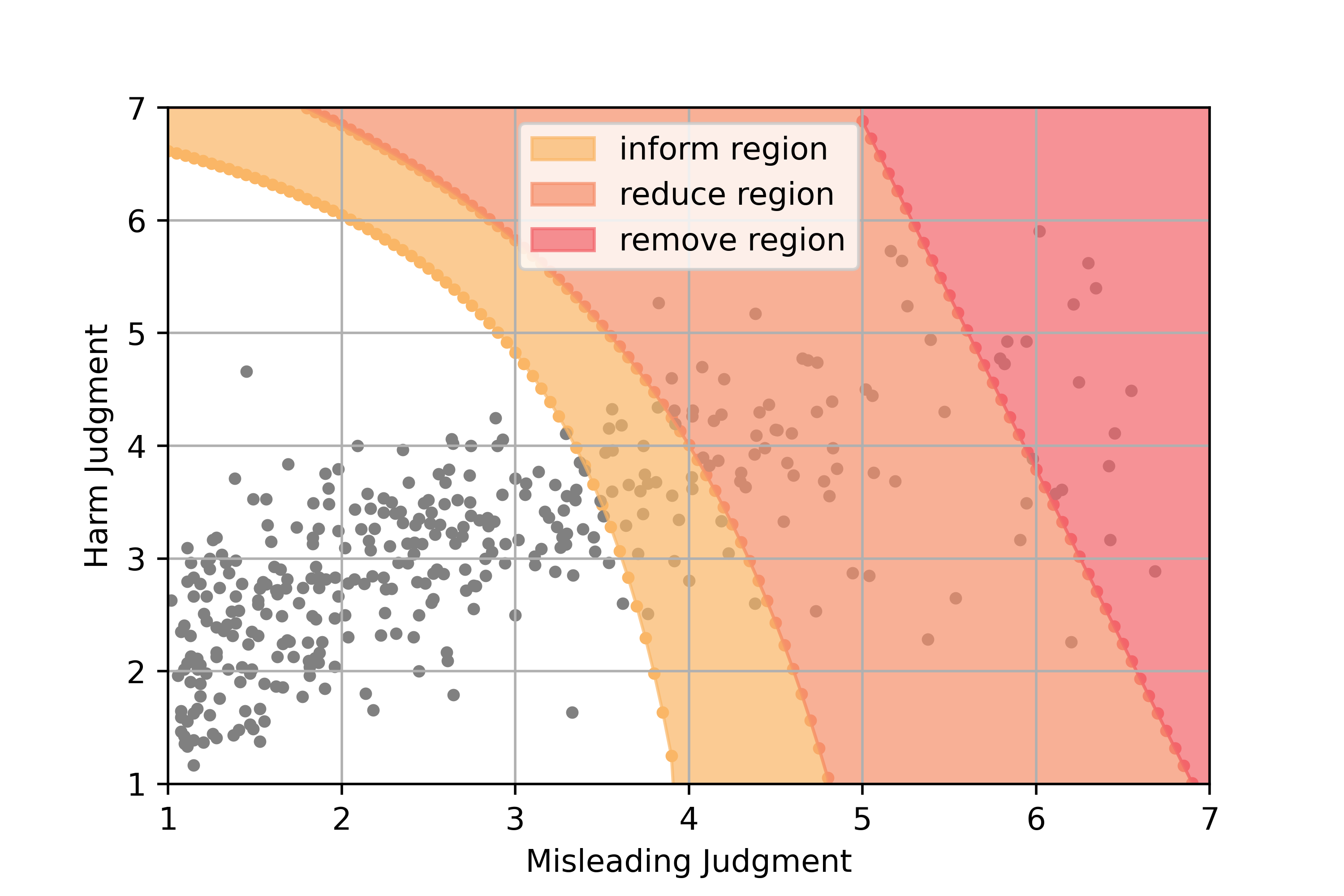}
    \caption{Visualizing the decision boundaries in terms of misleading and harm judgments for each action type}
    \label{fig:rq3:reg_eq}
\end{figure}

\begin{table}[!htbp] 
\centering 
\begin{tabular}{@{\extracolsep{5pt}}lccc}  \\[-1.8ex] \hline  \hline  \\[-1.8ex]  \\[-1.8ex] &  \multicolumn{1}{c}{Inform preference} &  \multicolumn{1}{c}{Reduce preference} &  \multicolumn{1}{c}{Remove preference}    \\   \\[-1.8ex] & (1) & (2) & (3)   \\  \hline   \\[-1.8ex]  Misleading judgment & 1.384$^{***}$ & 1.089$^{***}$ & 0.728$^{***}$   \\   & (0.055) & (0.055) & (0.021)   \\  Harm judgment & 0.803$^{***}$ & 0.682$^{***}$ & 0.236$^{***}$   \\   & (0.057) & (0.062) & (0.034)   \\  Misleading:Harm & -0.169$^{***}$ & -0.104$^{***}$ &  ------ \\   & (0.015) & (0.015) &   \\  Intercept & -5.575$^{***}$ & -5.422$^{***}$ & -5.259$^{***}$   \\   & (0.176) & (0.191) & (0.104)   \\  \hline   \\[-1.8ex]  AIC & 13082.630 & 12170.423 & 8030.254   \\  BIC & -191844.071 & -191825.658 & -191914.090   \\  Observations & 368 & 368 & 368   \\ \\  \hline  \hline   \\[-1.8ex]  \textit{Note:} &  \multicolumn{3}{r}{$^{*}$p$<$0.1; $^{**}$p$<$0.05; $^{***}$p$<$0.01}  \\  
\end{tabular} 
 \caption{Best performing regression models (based on AIC and BIC values) for predicting action preferences from misleading and harm judgments (RQ\ref{rq-proxy}).} 
 \label{tab:reg-proxy}
\end{table}

\rev{\textbf{Results:}} Table \ref{tab:reg-proxy} shows the regression coefficients for the best performing model for each action type. We find that models with both the predictors (misleading and harm judgments) perform better than the models with only one of them. For inform and reduce, the best performing model includes the interaction term as well, but not for remove. All the estimated models and their AIC-BIC values can be found in the Tables in Appendix~\ref{app:regression-results}. 

In order to interpret what each model has learned, Figure \ref{fig:rq3:reg_eq} plots the decision boundaries of the best fitting models for each action type in terms of misleading and harm judgments. That is, for each regression equation, we plot the values for misleading and harm judgments that lead to predicted action preferences of exactly 0.5; articles to the right of the curves are ones where the models predict that more than half of rater would prefer the corresponding action. 

If we were to rely on decision boundary of judgments to make the final action decisions, articles with misleading judgment greater than 3.95 would always be recommended for \textit{inform}. In addition, when the misleading judgment is greater than 4.85, \textit{reduce} would also be recommended as one of the actions. Higher harm scores can lead to the same (or a more stringent) action even when the misleading judgment is lower. For instance, an article with misleading and harm judgments of 4 and 3, respectively, would be recommended for \textit{inform}, while another article with scores of 4 and 5 would be recommended for \textit{reduce} as well. Articles with misleading judgment below 5 would never be recommended for \textit{remove}. When the misleading judgment is higher than 5, the type of action would still depend on the harm judgment. Furthermore, the \textit{reduce} and \textit{remove} decision curves do not intersect the harm-axis, which suggests that harm judgments alone may not be sufficient to recommend an action against an article. This makes sense because most people would not want to remove or reduce distribution of an article containing good information just because the topic was one where misinformation would be harmful. When an article has a harm score of 7, however, the decision rule would result in placing information label on the article. 
Also note that the decision boundary for the \textit{remove} action is a straight line, reflecting that the best fit model did not include an interaction term between the misleading and harm judgments.

Table \ref{tab:rq3:comparison} shows that decisions generated from the output of the misleadingness-harm regression models closely follow the decisions that would be made from taking the majority vote on the preference questions. By definition, decisions based on the majority's action preferences produce the fewest disagreements between individual preferences and the decisions. The level of disagreement would be only slightly higher if we used the predictions of the regression models based on aggregate judgments of misleading and harm (Table \ref{tab:rq3:comparison}, right side). 

\begin{figure}
    \centering
    \includegraphics[width=0.9\textwidth]{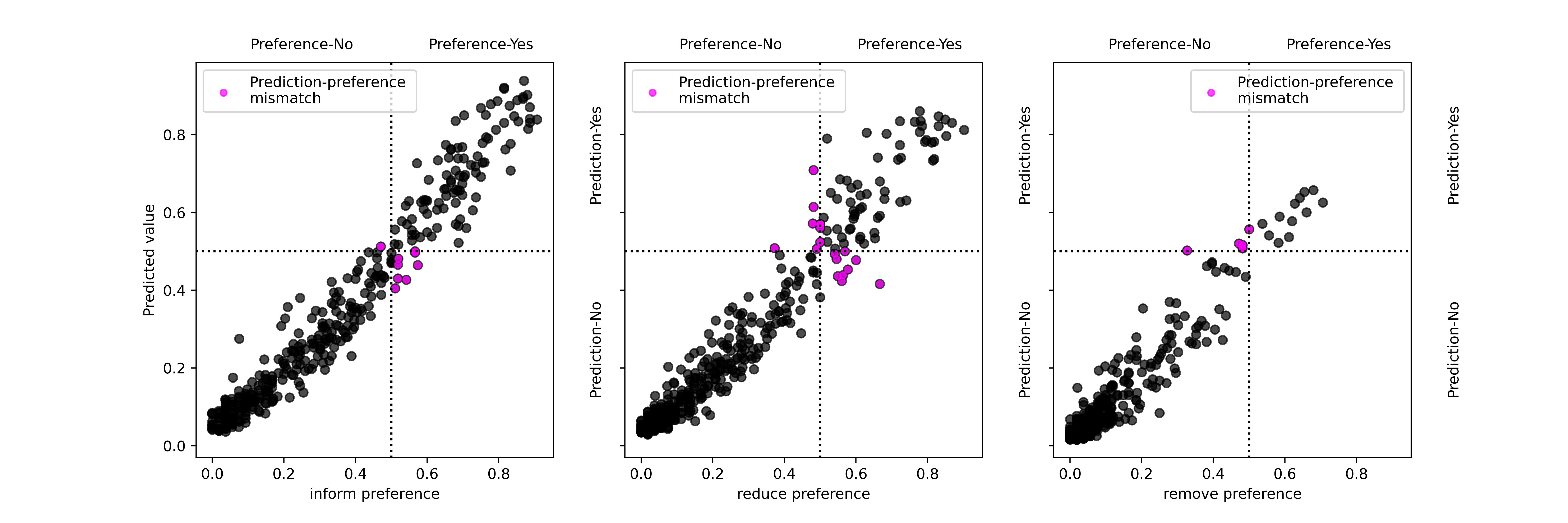}
    \caption{Comparison between actual preferences and the predicted preferences for each action type; the horizontal dotted line represents judgment-based decision boundary, the vertical dotted line represents preference-based decision boundary}
    \label{fig:rq3:cm}
\end{figure}

To understand why, despite some mismatches between the prediction-based decisions and preference-based decisions, the average levels of disagreement are largely similar, consider 
Figure~\ref{fig:rq3:cm}, which shows a confusion matrix-like representation. The vertical dotted line represents a decision-boundary based on expressed preferences of the majority (articles on the right are recommended for action; articles on the left are not) while the horizontal dotted line represents the decision boundary based on predictions from the regression model based on aggregate judgments (articles above the line are recommended for the action; articles below are not). We see that when there is a mismatch between the preference-based decision and judgment-based decision, the article is usually very close to the decision boundaries (color pink in Figure~\ref{fig:rq3:cm}), and thus either acting or not acting will both lead to half of raters disagreeing with the decision. 

We also analysed the set of articles that have a prediction-preference mismatch and found no obvious pattern in the topics or other features of the articles (a listing of those articles is in Appendix \ref{app:mismatch}). 

\begin{table}[]
\resizebox{0.7\textwidth}{!}{%
\begin{tabular}{l|c|c|c|c}
 & \multicolumn{2}{c|}{\textbf{Article set}}    &  \multicolumn{2}{c}{\textbf{Disagreement (\%age)}} \\ 
                  & \textbf{Preference-based} & \textbf{Prediction-based} &               \textbf{Preference-based} & \textbf{Prediction-based} \\ \hline
Inform            & 104              & 97 (-8/+1)                                                         & 23.57             & 23.76           \\ \hline
Reduce            & 70               & 70 (-9/+9)                                                           & 21.12             & 21.59             \\ \hline
Remove            & 12               & 17 (-0/+5)                                                            & 11.05            & 11.19             \\
\end{tabular}
}
\caption{Comparison between decisions based on aggregate preferences and decisions based on the regression model's output given the aggregate judgments of misleading and harm}
\label{tab:rq3:comparison}
\end{table}

\textbf{Generalizability of the models:}
We also tested whether the results we obtain are generalizable or not by evaluating our prediction performance on held-out test sets, using cross-validation. The cross-validation analysis found that results are robust and is provided as part of the Appendix (see \ref{app:pred}).

\section{Discussion}
\label{sec:discussion}

\subsection{You Can't Please Everyone}

Our results empirically show that there may not always be a clear consensus in terms of people's moderation preferences for individual content items. For instance, on 146 articles the preference for inform as the action was between 30-70\%. Similarly, there were 110 articles with reduce preference between 30-70\%. When remove was preferred by a majority, it was almost always a slim majority, with only one article achieving a supermajority of more than 70\%. On average, the level of disagreement with the majority preference ranges from 11.05\% for remove to 21.12\% for reduce and 23.57\% for inform. 


The lack of consensus cannot be explained entirely by ideological differences. We find differences in the preferences among ideologically-aligned users as well. For instance, if we consider only the liberal raters, on average about 18-20\% would disagree with the majority preference (of other liberals) for both reduce and inform actions (see Table~\ref{tab:mean-disapproval}).
Similarly, more than 20\% of conservatives would disagree with the majority preference among conservatives for both inform and reduce decisions. 

\rev{Lack of consensus matters because decisions that more people disagree with invite more complaints from the users. These complaints impact social media platforms by forcing them to spend additional resources to address these complaints and creating the risk of public outcry. When the contentiousness of platforms' decisions are revealed through public controversies \cite{gillespieCustodiansInternetPlatforms2018}, their decisions are often seen as arbitrary \cite{brennenTypesSourcesClaims}, or worse, biased \cite{borelFactCheckingWonUs2017}. Such controversies have reduced social media platforms' advertising revenue \cite{hern2020third} and resulted in a mass migration of users to other platforms \cite{schwedel2018did}. Our results, however, imply that substantial disagreement with any misinformation enforcement action or non-action is inevitable on many items, and will need to be managed rather than prevented.}


One implication is that the public discourse around platform enforcement practices needs to shift. The prevailing narrative is that the primary challenge in moderation decisions about misinformation is one of separating fact from fiction. Indeed, \rev{social media} platforms have partnered with independent fact-checking organizations to verify facts \cite{facebook-3pfc} while considerable emphasis has been put on scaling up the fact-checking process to handle the large volume of content circulating on social media platforms \cite{allen2020scaling,godel2021moderating,pintoFactCheckingCrowdsourcing2019,silvermanHelpingFactCheckersIdentify2019}. As we noted in the background section, these procedures result in a definitive outcome (in terms of whether and which action(s) to take) and fail to account for any differences in the public's preferences for these actions. 

Instead, legitimate differences of opinion may exist about what kinds of content are harmful to the public when they appear unsolicited in search results and social media feeds \cite{jiang2021understanding}. These differences may be driven by different judgments about how harmful the content would be if widely distributed \cite{bakerChallengesRespondingMisinformation2020}, different judgments about the harms that may come from reducing its distribution \cite{kozyreva2022free}, and individual differences in preferences about the tradeoffs between these harms. Like other consequential decisions where there are differences of opinion, they may need to be resolved through partially political processes. \rev{Social media} platforms should strive for a process that produces ``legitimate'' actions, outcomes that are broadly accepted even by people who do not agree with all of them. 





\subsection{Coping with Partisan Disagreement and Limiting the Tyranny of the Majority}


The debates in popular media \cite{kopitWhyBigTech2021, lemasterDebateIntensifiesFree} and some survey results \cite{MostAmericansThink2020} suggest that conservatives tend to be strong proponents of free speech and generally against any kind of censorship by social media platforms. However, by collecting action preferences on individual articles, we find no strong difference between \rev{the preferences of liberals and conservatives in our study with regard to how often actions should be taken overall.} 
\del{liberals' and conservatives' preferences with regard to how often each action needs be taken.}

We do find some differences in \textit{which} articles individuals think should be acted on: more conservatives would like to see action taken against articles from pro-liberal sources, and more liberals would like to see action taken against those from pro-conservative sources. Using these differing opinions to generate any kind of actionable insights runs the risk of increasing affective polarization \cite{iyengar2019origins}. Conservatives may believe that the judgments and preferences of liberal raters are causing reduced distribution of articles that conservatives approve of, and the effect may be exacerbated if those articles would primarily have been viewed by conservative readers. 

More generally, designing policies around the preferences of the majority runs the risk of creating a tyranny of the majority where outcomes are systematically bad for marginalized subgroups. Political theorists and practitioners have developed approaches to mitigating such risks. \rev{One such approach is to consider decision boundaries other than a simple majority (i.e., 50\%). For instance, social media platforms may require a supermajority (e.g., two-thirds majority) to justify an action \cite{leib2010deliberative,caluwaerts2014building}. As a result, enforcement actions that do not enjoy broad public support will not be implemented, and overall, fewer articles will be acted upon. Alternatively, a lower threshold may be set, creating more enforcement, as in the ``one-yes technique'' in a team of three \cite{kohler2017supporting} that gives power to take action to even a one-third minority. Under this approach, actions would be taken more frequently. Section \ref{decision-boundaries} in the Appendix analyzes how enforcement actions vary with different thresholds applied on the public's action preferences. It is important to note that a decision based on simple majority produces the least total disagreement among the public; either increasing or decreasing that threshold will increase the overall disagreement with the decision.} 

\rev{The impact of varying the threshold is also dependent on what constitutes as a ``default'' decision, as increasing the threshold will only leave more of the default decisions.} The current default of all major \rev{social media} platforms is non-action, allowing content to go viral unless an enforcement action curtails it \cite{gillespieCustodiansInternetPlatforms2018}. An alternative would be a default of ``friction'', with virality limited for all content unless an affirmative decision is taken to reduce the friction~\cite{diresta-friction, gillespie2022not}. In that case, the equation would be reversed as a higher threshold will afford more sensitivity to identifying potential harms. \rev{Overall, these thresholds and defaults uncover the tensions inherent in decisions made by social media platforms. Some may want to minimize the total number of user disagreements with their actions and inactions. Some may want to minimize the number of actions they take. Others may strive for more sensitivity to potential harms.}


Finally, setting a higher threshold may not always be sufficient to protect the interests of minority groups. For instance, if the primary risk to a minority group is over-moderation, more stringent criteria will mitigate the risk. If, however, the primary risk to minority groups is under-moderation (e.g., when misinformation circulating about a group is fomenting violence against them) then it might make sense to have lower thresholds. \rev{Other solutions to protect specific minority groups may require their representation in decision-making panels. We return to this in Section \ref{sec:future_work} as more work is needed to explore how minorities can be protected. }


\subsection{Eliciting Action Preferences vs. Misleadingness and Harm Judgments}


One of the challenges for eliciting action preferences is the need for raters to understand the space of potential enforcement actions. Our raters were required to pass a quiz that checked their understanding of the abstract action terms: inform, reduce, and remove. However, they could have answered those questions correctly by syntactic matching of words in the quiz questions and words in the definitions that we provided, without truly understanding them. One indication that our raters did understand them reasonably well is that their preferences implied a hierarchy of \rev{perceived} severity that makes intuitive sense, with remove \rev{perceived} as most severe, followed by reduce, and then informational labels \rev{perceived} as the least severe action. Moreover, this ordering largely held even when the interface reversed the order of presentation of the inform and reduce actions. 



We find that misleadingness judgments on their own were not sufficient to determine action preferences. 
Articles judged as extremely misleading or entirely false warranted action, but less severely misleading articles were also deemed actionable if the topic of the article was judged to be one where the public would be harmed by being misinformed (see the decision boundaries in Figure \ref{fig:rq3:reg_eq}). Furthermore, making decisions using both the misleadingness and harm judgments (and their corresponding thresholds) would yield action choices that would please almost as many raters as always choosing the majority preferred actions (see Table~\ref{tab:rq3:comparison}). Thus, it appears that in practice it may not be necessary to directly elicit action preferences at all. 

One caveat, however, is that we asked each rater to provide misleadingness and harm judgments before stating their action preferences. This could have encouraged them to make their action preferences correlate highly with their misleadingness and harm judgments. Future research could conduct a between-subjects test to see whether misleadingness and harm judgments from one group of raters can predict the action preferences of a different group of raters.

Setting aside that caveat, we speculate further that it may be sufficient to elicit just a single judgment from each rater, what we will call actionability. Raters could be asked to report the extent to which a content item is potentially harmful enough that some enforcement action should be taken. Different thresholds could be set on the mean actionability rating: above the lowest threshold, an inform action would be taken; above a somewhat higher threshold distribution would be reduced; at an even higher threshold, the item would be removed. Enforcement rules could even make a continuous mapping, where the extent to which the distribution of a content is reduced goes from 0\% to 100\% in line with the increase in the actionability ratings. 

\subsection{\rev{Limitations}}

\label{sec:limitations}
\rev{While our study provides novel insights about people's action preferences against potentially misleading content, the study design has several limitations. First, there might be biases in our rater pool that limit the generalizability of the results. We collected the raters' age, gender, and education qualification, and compared to the US population, our pool of raters skewed younger and more educated. It remains unclear whether a more representative sample of raters would provide similar results. Furthermore, in line with most prior literature on political ideology and misinformation \cite{pennycookFightingMisinformationSocial2019a,resnick2021survey}, our analysis on partisan differences is framed around the liberal-conservative dichotomy. We use a strict criteria to label conservatives and liberals (based on ideology and party affiliation) but other ideologies (e.g., authoritarian, libertarian) are not explicitly considered. We encourage future work to explore how misinformation action preferences differ along a multidimensional ideological lens.}

\rev{Another threat to generalizability may come from the set of articles we used in the study. While our dataset contained articles from many different sources, with a mix of misleading and non-misleading content, some of our findings may not extend beyond datasets with a similar distribution of sources. For instance, a dataset that is heavily biased toward articles from pro-conservative sources may show that liberal raters wanted more action than conservative raters. In the absence of any public information on the distribution of articles that circulate on social media platforms, we caution our readers against overgeneralizing our findings to other datasets and contexts.}

\rev{In our study, we solicited the raters' ``informed opinion'' by asking them to do some research and find corroborating evidence before providing their judgments and action preferences. However, these raters may be unaware of the larger implications of these enforcement actions or content moderation decisions in general. While we did quiz the participants on their understanding of individual actions, the quiz required them to only recall the instructions and not demonstrate any deeper understanding. Another study may be needed to compare and contrast the moderation preferences of those with demonstrated expertise or experience with content moderation, compared to our lay raters.}
 
\rev{The set of articles and raters in our study were limited to one country, the United States. Other than cultural and geopolitical issues, the public's action preference may also be impacted by their level of digital literacy and access to digital sources (e.g., via Google). Both these factors can impact the extent to which raters can find corroborating evidence online to make their judgments. A multi-nation cross-cultural study may be required to understand how these external factors impact the public's preferences for different enforcement actions.} 
 
\rev{Finally, how we framed the judgment and action preference questions may also impact the raters' responses. For instance, our question on ``misleadingness'' ranged from ``not misleading at all'' to ``false or extremely misleading''. In pilot testing, we found that focusing the question on misleadingness rather than truthfulness helped people think about the effect of the article as a whole, and labeling the extreme points as ``not misleading at all'' and ``false or extremely misleading'' was clear enough that workers were able to make judgments most of the time. However, it is possible that different design choices may result in different responses, and a more elaborate study design may be required to experiment with different task design choices and their impact on the raters. }

\subsection{\rev{Future work}}
\label{sec:future_work}

\rev{Future work should more rigorously examine how minority groups can be protected from any potential harms due to a consensus-based decision. As we noted earlier, setting different thresholds on the decision (such as the one-yes technique \cite{kohler2017supporting}) offer some protection but additional measures may also be required. One approach is to assure representation of at-risk groups in decision-making panels, a property known as ``descriptive representation''~\cite{pitkin1967concept, flanigan2021fair}. It may also be beneficial to have separate thresholds for subgroups, such as requiring approval of the majority of individual subgroups. For example, if an article is deemed harmful to a particular group, taking any action could require a majority of the at-risk group to prefer the action as well as at least 25\% of the overall population. Future work should investigate how such hybrid processes could be enacted and the effectiveness of these processes at protecting the interests of minority group. }

\rev{Another area for future work is to investigate how (and if) the public's perception of moderation processes is affected when they are informed about the disagreements concerning moderation decisions. For instance, prior work found that while social media users perceived expert panels to be more legitimate than lay juries, their perceptions were more strongly influenced by whether the decision aligned with their own preferences \cite{pan2022comparing}. Since these moderation decisions did not directly impact a given user, it is possible that these preferences are driven by a false consensus effect \cite{wojcieszak2009underlies}. Since our study demonstrates a frequent lack of consensus among individual users' action preferences, it remains to be seen how public's awareness of disagreements can impact their perception about the legitimacy of moderation processes.}

\rev{Given that differences will exist in individual users' moderation preferences, a third area to investigate is how to account for these differences at a scale that social media platforms operate. For example, can we collect users' preferences in near real-time as content is getting circulated such that these preferences represent a sufficiently large and diverse group of users? Alternatively, the scope may be limited to only collect users' preferences for auditing social media platform's decisions \cite{eslami2019user} or for reviewing appeals about decisions that have already been made. Furthermore, if platforms are using alternate processes (e.g., juries or algorithms) to make decisions, future work should investigate whether or how these processes can evolve to accommodate the differences in users' preferences (see \cite{gordon2022jury} for example).}

\section{Conclusion}


\rev{Social media} platforms will not please everyone all the time no matter which actions they pursue against misinformation. Instead, our results indicate that even on many individual articles, public opinions are split. Given the variation in \rev{individuals'} action preferences, it may be helpful to reframe expectation for platforms. Instead of expecting them to produce \textit{correct} decisions, the public should expect platforms to make \textit{legitimate} decisions where even people who disagree with particular decisions can agree the process and outcome were appropriate.

One source of legitimacy can come from following transparent procedures based on pre-announced policies that map from quasi-objective attributes to enforcement actions. \rev{Social media} platforms largely try to follow this approach today. The legitimacy of this approach can be enhanced by appealing to independent authorities to set policies to cover challenging cases, as with Facebook's independent \del{o}\rev{O}versight \del{b}\rev{B}oard~\cite{Facebook-Oversight-Board}.

This paper introduces another potential way to produce legitimate decisions -- by accumulating a large set of cases where an informed public expresses their preferences for moderation actions, and then crafting policy rules that try to predict those preferences. This approach is analogous to Oliver Wendell Holmes\rev{'} conception of how the legal system works~\cite{holmes1997path}. He argued that decision rules are prophecies about how courts will decide cases, not definitions of what the correct outcomes are. Many details of procedures would need to be worked out in order to craft misinformation moderation decisions around the collective judgments of citizen panels on particular items. 


Initial evidence suggests that the approach is promising. It does appear to be possible to craft decision rules that can predict action preferences based on judgments about properties of the content. Majority \rev{action} preferences were not entirely predictable from misleadingness judgments alone, but they were predictable from a combination of misleadingness and harm judgments. 






\bibliographystyle{ACM-Reference-Format}
\bibliography{main}

\appendix
\section{Appendix}
\label{sec:appendix}
\revminor{\subsection{Screenshots of the rating interface used for the study}}
\label{sec:screenshots}

\begin{figure}[H]
    \centering
    \fbox{\includegraphics[width=0.7\textwidth]{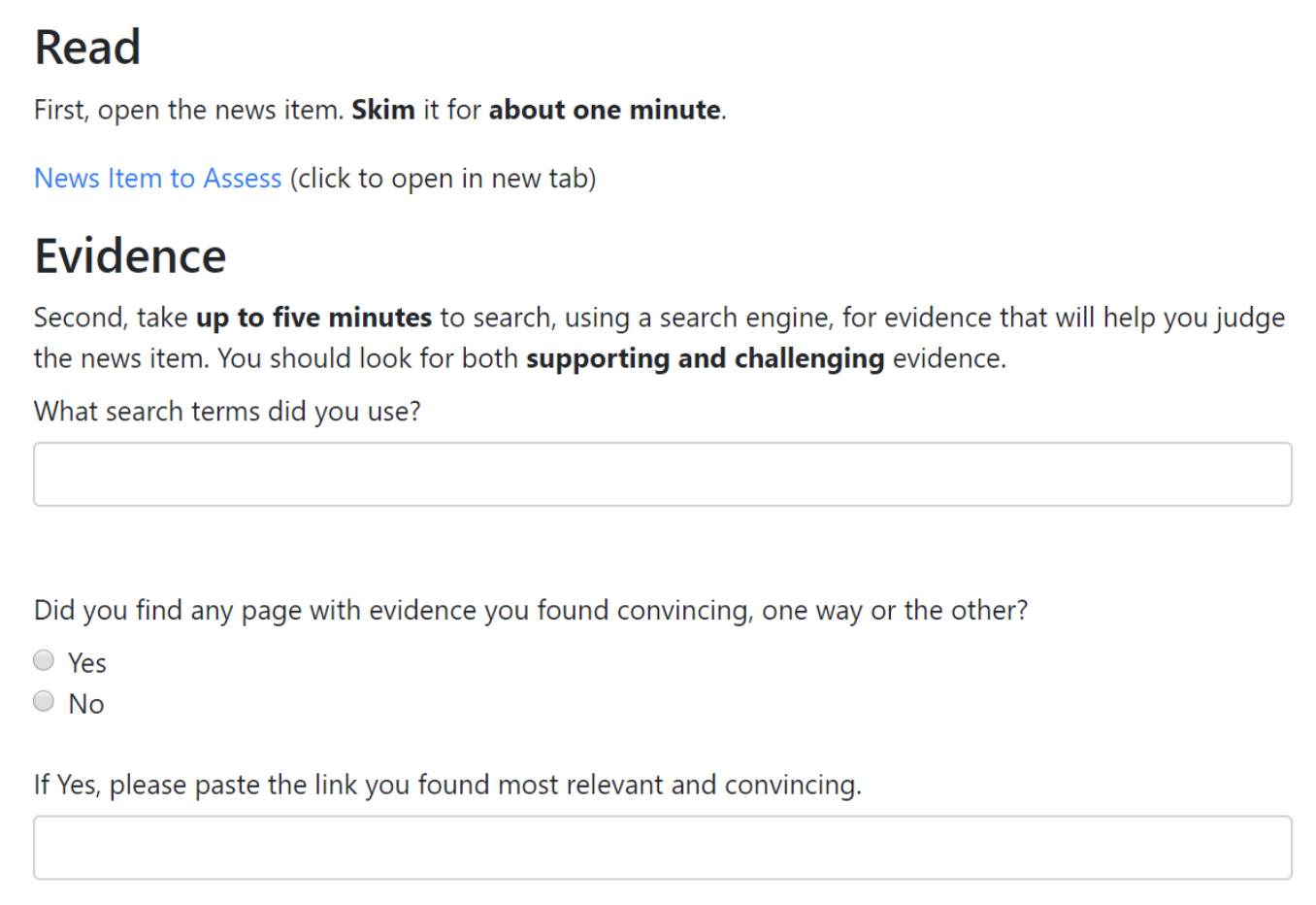}}
    \caption{Presenting the news article and asking for evidence}
    \label{fig:screen1}
\end{figure}

\begin{figure}[H]
    \centering
    \fbox{\includegraphics[width=0.7\textwidth]{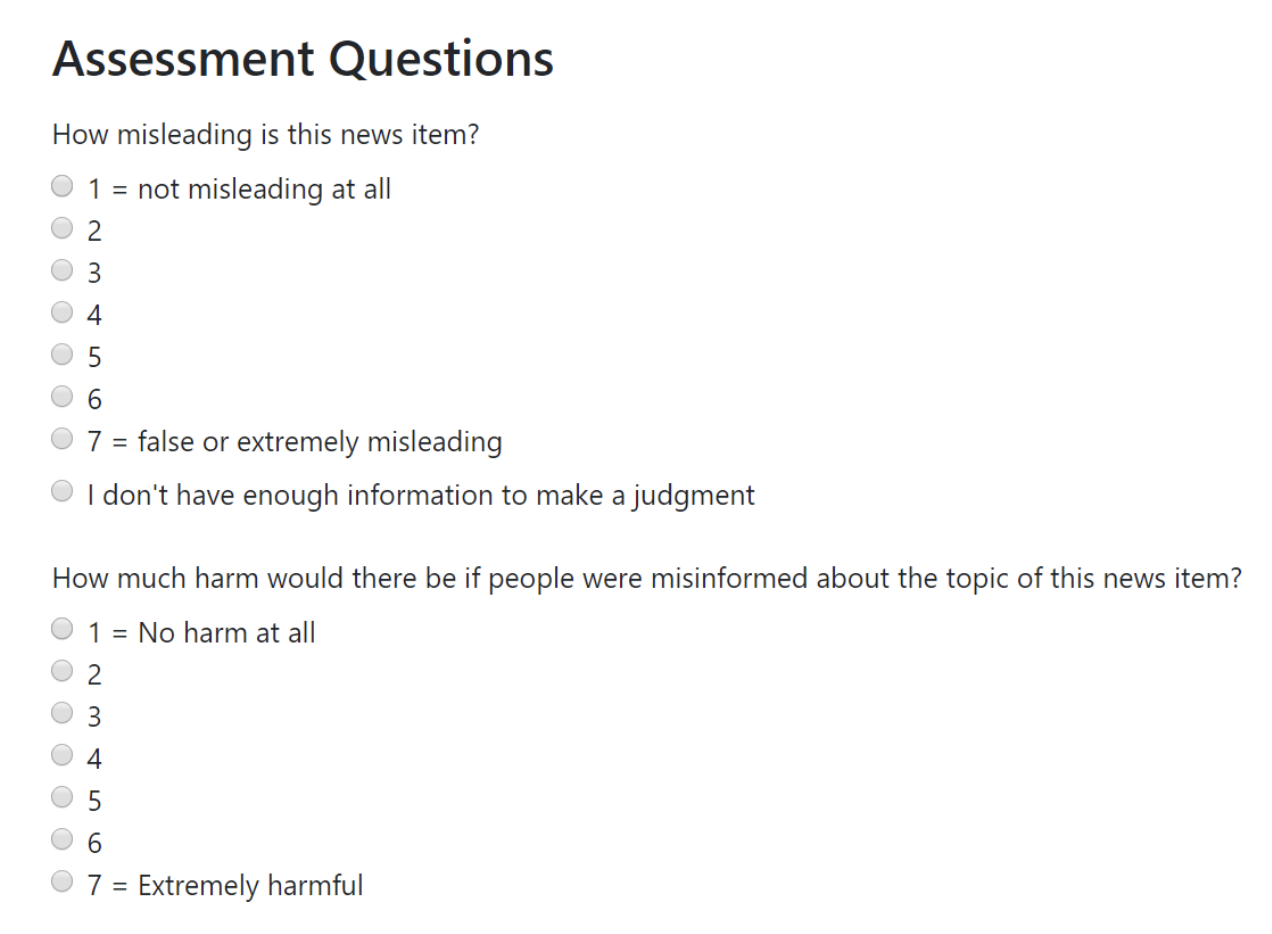}}
    \caption{Asking for judgments}
    \label{fig:screen2}
\end{figure}

\begin{figure}[H]
    \centering
    \fbox{\includegraphics[width=0.8\textwidth]{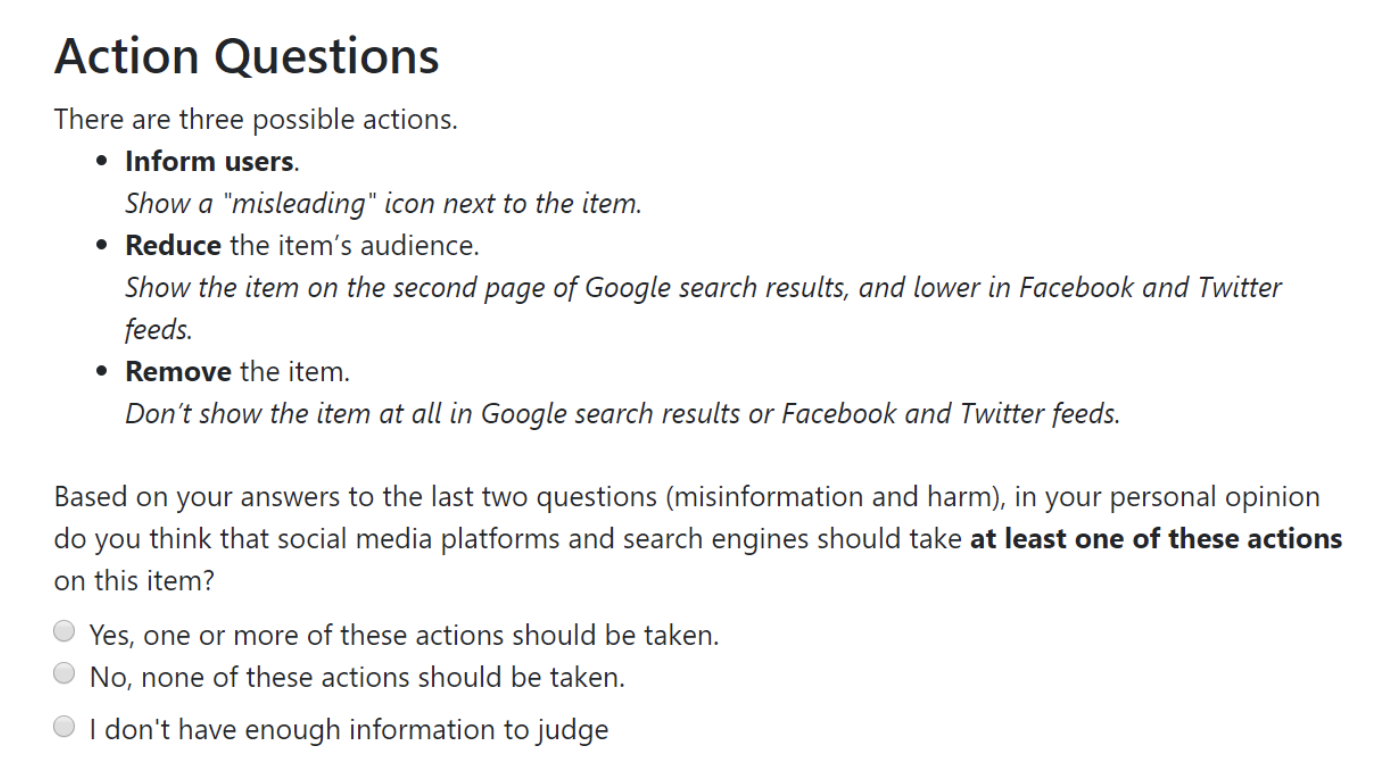}}
    \caption{Explaining action types}
    \label{fig:screen3}
\end{figure}

\begin{figure}[H]
    \centering
    \fbox{\includegraphics[width=0.7\textwidth]{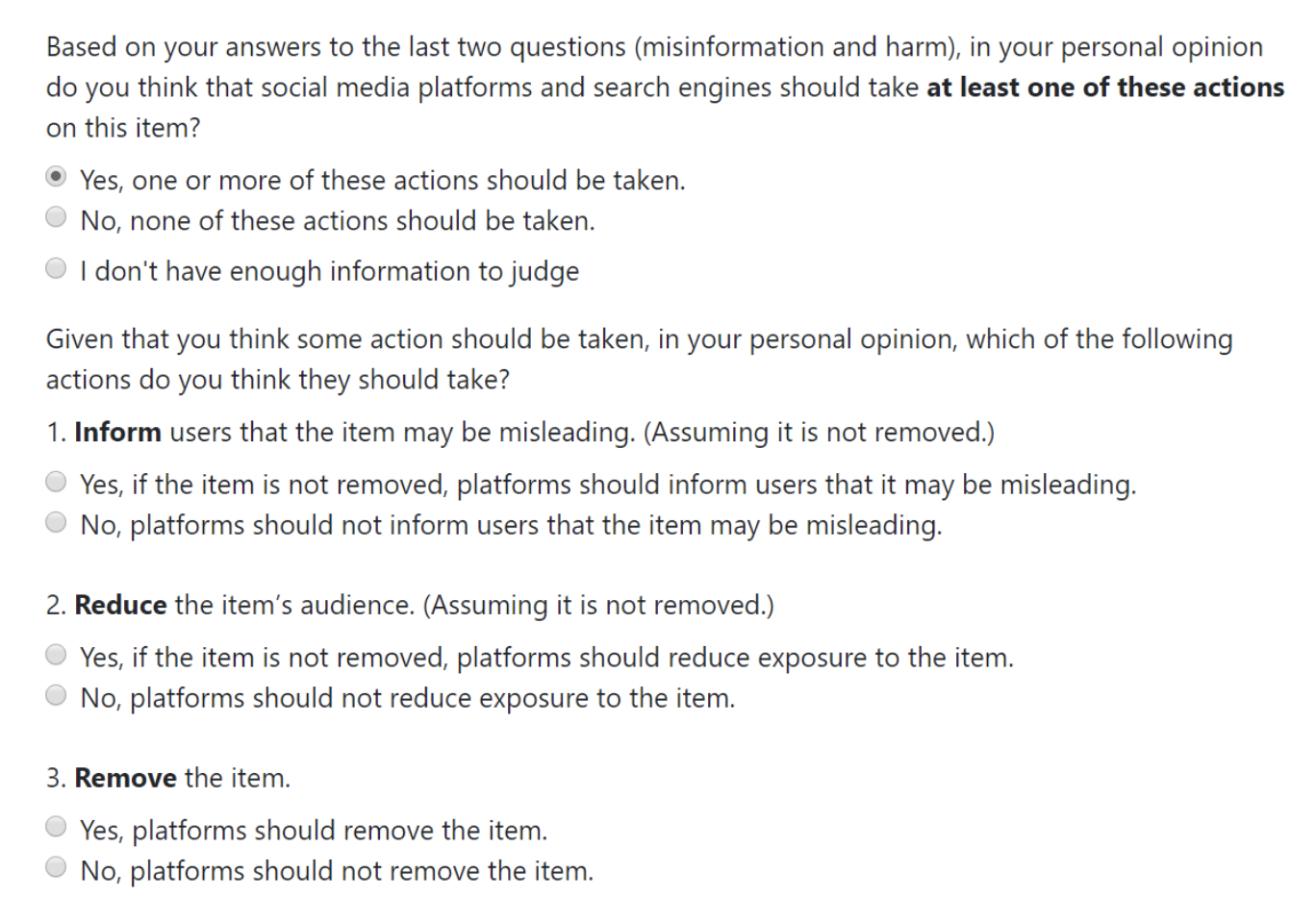}}
    \caption{Asking for action preferences}
    \label{fig:screen4}
\end{figure}

\rev{\subsection{Distribution of Journalist Ratings for Articles}}
\label{sec:journalist_dist}

\begin{figure}[H]
    \centering
    \includegraphics[width=0.6\textwidth]{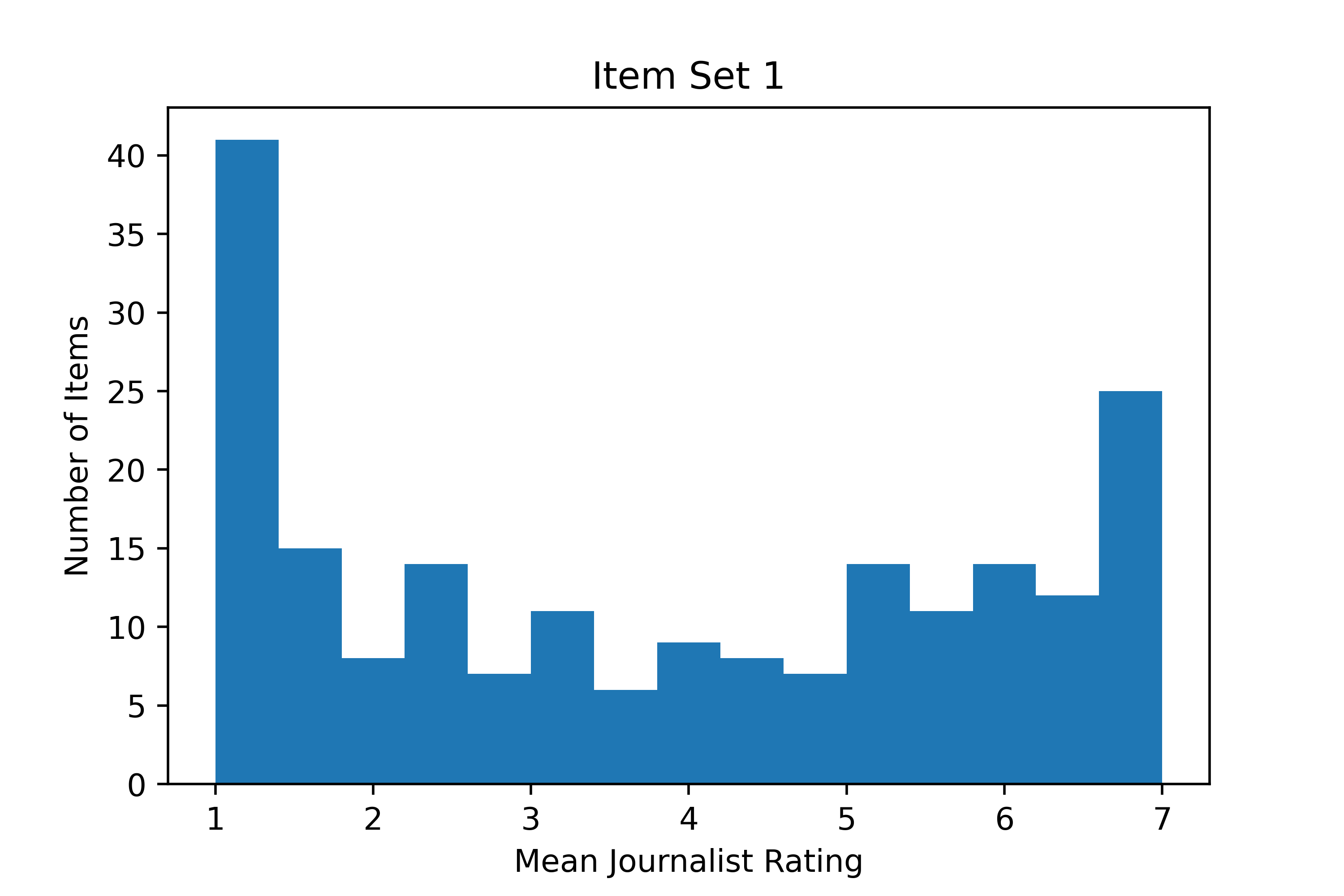}
    \caption{Distribution of the journalists' mean misleading judgments on the first collection of articles (provided by Facebook).}
    \label{fig:journalists-mit}
\end{figure}

\begin{figure}[H]
    \centering
    \includegraphics[width=0.6\textwidth]{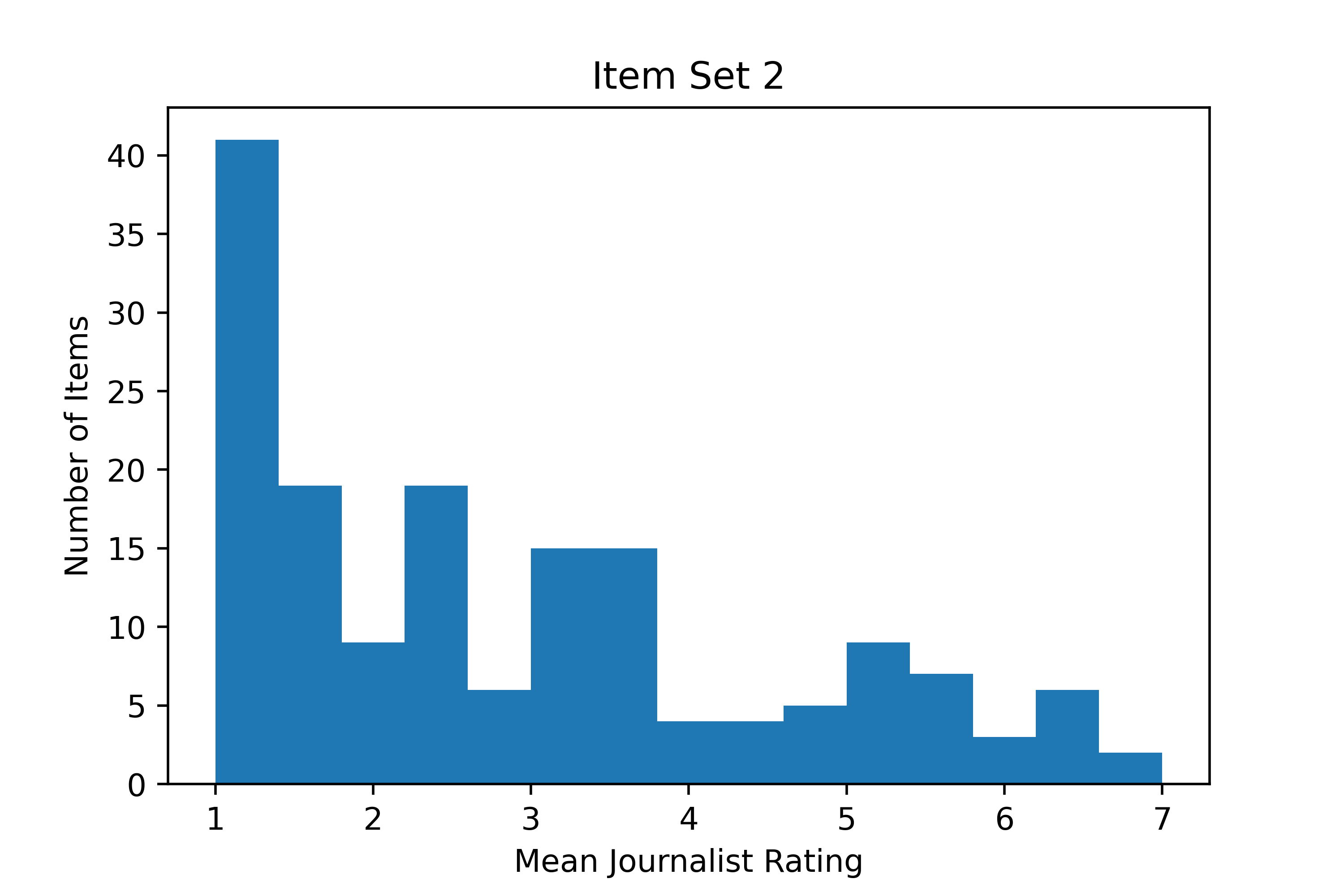}
    \caption{Distribution of the journalists' mean misleading judgments for the second collection of articles (provided by \cite{godel2021moderating})}
    \label{fig:journalists-nyu}
\end{figure}

\subsection{Articles recommended for remove}
\label{app:remove}
\begin{table}[H]
    \centering
    \resizebox{\textwidth}{!}{
\begin{tabular}{p{0.7\textwidth}lrrrrr}\toprule                                                                                                                                                           url & source bias &  misleading judgment &  harm judgment &  inform preference &  reduce preference &  remove preference \\\midrule                                                            https://neonnettle.com/news/4335-obama-at-bilderberg-the-us-must-surrender-to-the-new-world-order- &     no\_bias &                 6.35 &           5.40 &               0.76 &               0.76 &               0.65 \\                                                          http://www.breakingnews247.net/5b872f00c5639/massive-alligator-found-in-browns-mills-new-jersey.html &     no\_bias &                 6.43 &           3.17 &               0.81 &               0.72 &               0.56 \\                                                      https://worldnewsdailyreport.com/cops-beat-up-teen-after-bank-teller-mistakes-his-erection-for-a-pistol/ &     no\_bias &                 5.83 &           4.93 &               0.81 &               0.80 &               0.61 \\                                                            https://worldnewsdailyreport.com/woman-arrested-for-training-squirrels-to-attack-her-ex-boyfriend/ &     no\_bias &                 6.69 &           2.89 &               0.87 &               0.78 &               0.54 \\                                                                              http://www.55meals.com/did-you-know-your-energy-drinks-contain-bull-urine-semen/ &     missing &                 6.25 &           4.57 &               0.89 &               0.83 &               0.58 \\                                                               https://worldnewsdailyreport.com/morgue-worker-arrested-after-giving-birth-to-a-dead-mans-baby/ &     no\_bias &                 6.45 &           4.11 &               0.87 &               0.85 &               0.66 \\                                                     https://www.infowars.com/nearly-200-people-arrested-across-australia-for-deliberately-starting-bushfires/ &        cons &                 5.82 &           4.73 &               0.85 &               0.82 &               0.58 \\https://worldtruth.tv/the-coronavirus-was-engineered-by-scientists-in-a-lab-using-well-documented-genetic-engineering-vectors-that-leave-behind-a-fingerprint/ &     no\_bias &                 6.30 &           5.62 &               0.89 &               0.87 &               0.68 \\                                                https://worldnewsdailyreport.com/teen-on-female-viagra-crashes-into-building-while-masturbating-to-gear-shift/ &     no\_bias &                 6.42 &           3.82 &               0.88 &               0.78 &               0.62 \\                                                      http://healthimpactnews.com/2011/dr-russell-blaylock-warns-dont-get-the-flu-shot-it-promotes-alzheimers/ &     no\_bias &                 6.02 &           5.90 &               0.88 &               0.90 &               0.71 \\                      https://endoftheageheadlines.wordpress.com/2018/10/24/deep-state-sending-explosive-packages-to-themselves-in-hopes-of-stopping-red-wave/ &     missing &                 6.22 &           5.25 &               0.84 &               0.78 &               0.63 \\                                                    https://worldnewsdailyreport.com/pregnant-teen-seeks-13-paternity-tests-after-gangbang-with-football-team/ &     no\_bias &                 6.55 &           4.49 &               0.87 &               0.83 &               0.64 \\\bottomrule\end{tabular}
}
\end{table}

\subsection{Regression results}
\label{app:regression-results}

\begin{table}[H] \centering  \begin{tabular}{@{\extracolsep{5pt}}lcccc} \\[-1.8ex]\hline \hline \\[-1.8ex] & \multicolumn{4}{c}{\textit{Dependent variable: Inform preference}} \\ \\[-1.8ex] & \multicolumn{1}{c}{Misleading only} & \multicolumn{1}{c}{Harm only} & \multicolumn{1}{c}{Both} & \multicolumn{1}{c}{Interaction}  \\ \\[-1.8ex] & (1) & (2) & (3) & (4) \\ \hline \\[-1.8ex]  Misleading judgment & 0.954$^{***}$ & & 0.846$^{***}$ & 1.316$^{***}$ \\   & (0.018) & & (0.023) & (0.067) \\  Harm judgment & & 1.054$^{***}$ & 0.246$^{***}$ & 0.649$^{***}$ \\   & & (0.025) & (0.033) & (0.064) \\  Misleading:Harm & & & & -0.133$^{***}$ \\   & & & & (0.017) \\  Intercept & -3.325$^{***}$ & -4.080$^{***}$ & -3.831$^{***}$ & -5.170$^{***}$ \\   & (0.057) & (0.087) & (0.091) & (0.206) \\ \hline \\[-1.8ex]  AIC & 8893.05 & 10655.89 & 8839.80 & 8786.07 \\  BIC & -122673.37 & -120910.53 & -122719.15 & -122765.41 \\  Observations & 368 & 368 & 368 & 368 \\ \hline \hline \\[-1.8ex] \textit{Note:} & \multicolumn{4}{r}{$^{*}$p$<$0.1; $^{**}$p$<$0.05; $^{***}$p$<$0.01} \\ \end{tabular} \label{tab:reg1}
\caption{Predicting preferences for inform action using misleading and potential harm scores}
\end{table}

\begin{table}[H] \centering  \begin{tabular}{@{\extracolsep{5pt}}lcccc} \\[-1.8ex]\hline \hline \\[-1.8ex] & \multicolumn{4}{c}{\textit{Dependent variable: Reduce preference}} \\ \\[-1.8ex] & \multicolumn{1}{c}{Misleading only} & \multicolumn{1}{c}{Harm only} & \multicolumn{1}{c}{Both} & \multicolumn{1}{c}{Interaction}  \\ \\[-1.8ex] & (1) & (2) & (3) & (4) \\ \hline \\[-1.8ex]  Misleading judgment & 0.871$^{***}$ & & 0.750$^{***}$ & 1.013$^{***}$ \\   & (0.017) & & (0.021) & (0.064) \\  Harm judgment & & 1.041$^{***}$ & 0.292$^{***}$ & 0.548$^{***}$ \\   & & (0.026) & (0.033) & (0.068) \\  Misleading:Harm & & & & -0.075$^{***}$ \\   & & & & (0.017) \\  Intercept & -3.555$^{***}$ & -4.444$^{***}$ & -4.183$^{***}$ & -5.025$^{***}$ \\   & (0.059) & (0.093) & (0.096) & (0.221) \\ \hline \\[-1.8ex]  AIC & 8365.07 & 9757.62 & 8289.63 & 8272.79 \\  BIC & -122631.30 & -121238.74 & -122699.26 & -122708.63 \\  Observations & 368 & 368 & 368 & 368 \\ \hline \hline \\[-1.8ex] \textit{Note:} & \multicolumn{4}{r}{$^{*}$p$<$0.1; $^{**}$p$<$0.05; $^{***}$p$<$0.01} \\ \end{tabular} \label{tab:reg2} 
\caption{Predicting preferences for reduce action using misleading and potential harm scores}
\end{table}

\begin{table}[H] \centering \begin{tabular}{@{\extracolsep{5pt}}lcccc} \\[-1.8ex]\hline \hline \\[-1.8ex] & \multicolumn{4}{c}{\textit{Dependent variable: Remove preference}} \\   \\[-1.8ex] & \multicolumn{1}{c}{Misleading only} & \multicolumn{1}{c}{Harm only} & \multicolumn{1}{c}{Both} & \multicolumn{1}{c}{Interaction}  \\ \\[-1.8ex] & (1) & (2) & (3) & (4) \\ \hline \\[-1.8ex]  Misleading judgment & 0.805$^{***}$ & & 0.727$^{***}$ & 0.670$^{***}$ \\   & (0.019) & & (0.024) & (0.070) \\  Harm judgment & & 0.969$^{***}$ & 0.214$^{***}$ & 0.143$^{}$ \\   & & (0.031) & (0.038) & (0.091) \\  Misleading:Harm & & & & 0.016$^{}$ \\   & & & & (0.019) \\  Intercept & -4.656$^{***}$ & -5.348$^{***}$ & -5.150$^{***}$ & -4.918$^{***}$ \\   & (0.081) & (0.119) & (0.123) & (0.297) \\ \hline \\[-1.8ex]  AIC & 5609.69 & 6538.64 & 5580.18 & 5581.46 \\  BIC & -122761.78 & -121832.83 & -122783.81 & -122775.07 \\  Observations & 368 & 368 & 368 & 368 \\   \hline \hline \\[-1.8ex] \textit{Note:} & \multicolumn{4}{r}{$^{*}$p$<$0.1; $^{**}$p$<$0.05; $^{***}$p$<$0.01} \\ \end{tabular} \label{tab:reg3} 
\caption{Predicting preferences for remove action using misleading and potential harm scores}
\end{table}

\subsection{Action-Judgment Mismatch}
\label{app:mismatch}
\begin{table}[H]
    \centering
   \resizebox{\textwidth}{!}{%
   \begin{tabular}{p{0.7\textwidth}lrrrr} \toprule                                                                      url & source bias &  misleading judgment &  harm judgment &  inform preference &  inform prediction \\\midrule                                  https://www.collective-evolution.com/2020/01/22/another-supposedly-authentic-photo-of-a-ufo-the-story-behind-it/ &     no\_bias &                 3.76 &           2.51 &               0.47 &               0.51 \\                  \midrule                                                    https://www.palmerreport.com/analysis/retweeting-bizarre-fake-account/23745/ &         lib &                 3.15 &           3.09 &               0.51 &               0.40 \\                  https://washingtonpress.com/2018/10/28/pittsburgh-jewish-leaders-just-banned-trump-from-their-city-until-he-meets-their-demands/ &         lib &                 3.33 &           2.85 &               0.52 &               0.43 \\                                       https://www.lifezette.com/2018/10/kavanaugh-turned-down-scads-of-gofundme-dollars-blasey-ford-hits-paydirt/ &        cons &                 3.39 &           3.26 &               0.52 &               0.47 \\                  https://fellowshipoftheminds.com/pentagon-bans-bible-verses-on-dog-tags-while-pres-trump-upholds-right-to-pray-in-public-schools &     no\_bias &                 3.54 &           2.96 &               0.52 &               0.48 \\                                 https://friendsforsyria.com/2019/12/04/ukrainian-neo-nazis-help-out-at-hong-kong-riots-pan-democrats-defend-them/ &     missing &                 3.06 &           3.67 &               0.54 &               0.43 \\                                                  https://www.breitbart.com/big-journalism/2018/08/18/cnn-accused-intimidating-paul-manafort-jury/ &        cons &                 3.49 &           3.51 &               0.57 &               0.50 \\                                      https://www.naturalnews.com/2020-01-20-san-fran-democrat-tyrants-taxing-landlords-leaving-stores-vacant.html &        cons &                 3.51 &           3.38 &               0.57 &               0.50 \\https://www.movieguide.org/news-articles/netflix-animated-series-dedicates-an-entire-episode-to-promote-planned-parenthood-and-killing-babies.html &     missing &                 3.30 &           3.56 &               0.57 &               0.46 \\\bottomrule\end{tabular}'
   }
    \caption{Inform mismatch}

\end{table}

\begin{table}[H]
    \centering
   \resizebox{\textwidth}{!}{%
   \begin{tabular}{p{0.7\textwidth}lrrrr} \toprule                                                                              url & source bias &  misleading judgment &  harm judgment &  remove preference &  remove\_prediction \\\midrule                         https://babylonbee.com/news/bill-clinton-allegations-of-sexual-misconduct-should-disqualify-a-man-from-public-office &     no\_bias &                 5.98 &           3.88 &               0.33 &               0.50 \\              https://americanmilitarynews.com/2018/08/china-hacked-hillary-clintons-email-server-and-took-nearly-all-her-emails-report-says/ &     no\_bias &                 5.79 &           4.77 &               0.47 &               0.52 \\                                                                       http://www.worldstarhiphop.com/videos/video.php?v=wshhddDiUTw9SDG7wvd7 &     no\_bias &                 6.15 &           3.61 &               0.48 &               0.52 \\https://www.nsfnews.com/5b8ea8e312074/jackson-man-arrested-for-hacking-a-college-computer-and-returning-all-funds-to-students-since-2010.html &     missing &                 6.11 &           3.57 &               0.48 &               0.51 \\                                                              https://conservativedailypost.com/savage-claims-ford-deeply-tied-to-deep-state/ &        cons &                 5.94 &           4.93 &               0.50 &               0.56 \\\bottomrule\end{tabular}
   }
    \caption{Remove mismatch}
\end{table}

\begin{table}[H]
    \centering
   \resizebox{\textwidth}{!}{%
   \begin{tabular}{p{0.7\textwidth}lrrrr} \toprule                                                                         url & source bias &  misleading judgment &  harm judgment &  reduce preference &  reduce prediction \\\midrule                                                  https://www.dailywire.com/news/37685/epa-greenhouse-gas-emissions-dropped-nearly-3-joseph-curl &        cons &                 4.12 &           3.82 &               0.37 &               0.51 \\                                                                                            https://patriotjournal.org/video-train-south-border/ &        cons &                 4.44 &           3.98 &               0.48 &               0.57 \\                  http://coolcatapproves.com/funny/australia-doesnt-exist-and-people-who-live-there-are-actors-paid-by-nasa-flat-earthers-claim/ &     missing &                 4.78 &           3.69 &               0.48 &               0.61 \\                                                            https://babylonbee.com/news/joel-osteen-launches-line-pastoral-wear-sheeps-clothing/ &     no\_bias &                 5.54 &           2.65 &               0.48 &               0.71 \\                  https://www.teaparty.org/breaking-ukrainian-official-reveals-six-criminal-cases-opened-in-ukraine-involving-the-bidens-420208/ &        cons &                 4.08 &           3.90 &               0.49 &               0.51 \\ https://www.breitbart.com/politics/2018/11/01/orourke-campaign-exposed-in-undercover-video-for-assisting-honduran-migrants-nobody-needs-to-know/ &        cons &                 4.39 &           4.09 &               0.50 &               0.57 \\                                                                http://alexschadenberg.blogspot.com/2018/10/sick-kids-hospital-toronto-will.html &     missing &                 4.20 &           4.59 &               0.50 &               0.57 \\                                                                                                      https://realfarmacy.com/surgeon-mammogram/ &     no\_bias &                 3.83 &           5.27 &               0.50 &               0.56 \\                                  https://www.concealedcarry.com/news/armed-citizens-are-successful-95-of-the-time-at-active-shooter-events-fbi/ &     missing &                 4.02 &           4.31 &               0.50 &               0.52 \\              
\midrule 
https://www.dailywire.com/news/38153/breaking-voter-fraud-allegedly-found-deep-blue-ryan-saavedra &        cons &                 3.82 &           4.34 &               0.54 &               0.49 \\                                                                  https://www.zeptha.com/cotton-swab-soaked-in-alcohol-and-placed-in-your-navel/ &     missing &                 4.23 &           3.05 &               0.55 &               0.48 \\                              https://www.breitbart.com/border/2018/10/30/armed-migrants-in-caravan-opened-fire-on-mexican-cops-say-authorities/ &        cons &                 3.76 &           3.67 &               0.55 &               0.44 \\                            https://www.palmerreport.com/analysis/trumps-sham-acquittal-is-already-blowing-up-in-senate-republicans-faces/24893/ &         lib &                 3.72 &           3.60 &               0.56 &               0.42 \\             https://americanmilitarynews.com/2018/10/guatemala-captured-100-isis-terrorists-president-reveals-ahead-of-migrant-caravan-arrival/ &     no\_bias &                 3.75 &           3.75 &               0.56 &               0.44 \\                                         https://legalinsurrection.com/2018/09/maxine-waters-suggests-knocking-off-trump-then-going-after-pence/ &        cons &                 3.92 &           4.20 &               0.57 &               0.50 \\                                                              https://www.healthy-holistic-living.com/instant-noodles-inflammation-dementia.html &     no\_bias &                 3.90 &           3.56 &               0.58 &               0.45 \\                                                                            http://www.higherperspectives.com/one-glass-red-wine-1577145867.html &     no\_bias &                 4.02 &           3.62 &               0.60 &               0.48 \\                                                                    https://www.palmerreport.com/analysis/annoymous-rudy-giuliani-berserk/23040/ &         lib &                 3.92 &           2.98 &               0.67 &               0.42 \\\bottomrule\end{tabular}
   }
   \caption{Reduce mismatch}
\end{table}

\subsection{Generalizability of Results for RQ\ref{rq-proxy}}
\label{app:pred}

We tested whether the results we obtain for RQ\ref{rq-proxy} are generalizable or not by evaluating our prediction performance on held-out test sets, using cross-validation. We train our models over 10 iterations, each time using a different partial dataset (80\% -- 294 articles) and report the prediction performance on the corresponding held-out test set (20\% -- 74 articles). 

The cross-validation analysis shows that the results are robust. Table \ref{tab:rq3:cv} shows the prediction performance in terms of the level of disagreement on only the held-out test-set articles, averaged over ten iterations. Even on held-out articles, the level of disagreement with the action selected from the prediction of the regression model from the misinformation and action judgments produces an average level of individual rater disagreement that is only marginally higher than the minimum that could be achieved by always acting on the majority action preference.

Additionally, we also report two standard metrics for evaluating predictions -- namely, the F1 score and the Jensen Shannon (JS) distance. We use F1 score to measure the classification accuracy, i.e., whether the predicted decision (action or no action) matches with the preference-based decision. We also use Jensen Shannon (JS) distance to measure the distance between the underlying distributions, i.e., actual preferences and predicted preferences, without considering the decision outcome. The average result from 10 iterations are reported in Table \ref{app:tab:rq3:cv}. We find the F1 score and the JS distance to be largely consistent over these iterations. The high values of F1 score and low values of JS distance helps establish that our models are generalizable and can be used to predict action preferences on completely new articles as well. 

\begin{table}[]
\resizebox{0.6\textwidth}{!}{%
\begin{tabular}{l|c|c}

       & \textbf{Preference-based} & \textbf{Prediction-based} \\ \hline
Inform & 23.28 (SD: 1.72)     & 23.60 (SD: 1.86)                   \\ \hline
Reduce & 21.44 (SD: 1.57)    & 21.89 (SD: 1.57)                  \\ \hline
Remove & 11.7 (SD: 0.98)     & 11.86 (SD: 1.02)                  \\ 
\end{tabular}
}
\caption{Prediction performance in terms of the level of disagreement (\%age users) on a held-out test set (74 articles) over 10 iterations}
\label{tab:rq3:cv}
\end{table}

\begin{table}[]
\resizebox{0.6\textwidth}{!}{%
\begin{tabular}{|l|l|l|}
\hline
       & F1 Score & Jensen Shannon Distance \\ \hline
Inform & 0.92 (SD: 0.03)     & 0.08 (SD: 0.009)                   \\ \hline
Reduce & 0.89 (SD: 0.05)    & 0.11 (SD: 0.009)                  \\ \hline
Remove & 0.89 (SD: 0.10)     & 0.16 (SD: 0.019)                  \\ \hline
\end{tabular}
}
\caption{Prediction performance using F1 score and JS distance on a held out test set (74 articles) over 10 iterations}
\label{app:tab:rq3:cv}
\end{table}

\label{decision-boundaries}
\subsection{\rev{Exploring other Decision Boundaries}}

\begin{table}[]
\begin{tabular}{|l|l|l|l|}
\hline
\multicolumn{4}{|c|}{Articles recommended for action}                                                                \\ \hline
Decision boundary               & Inform & Reduce & Remove \\ \hline
25\%           & 203    & 157    & 56     \\ \hline
33.33\%        & 169    & 121    & 37     \\ \hline
\textbf{50\% (default)} & 104    & 70     & 12     \\ \hline
66.66\%        & 63     & 31     & 3      \\ \hline
75\%           & 28     & 17     & 0      \\ \hline
\end{tabular}
\caption{No. of articles recommended for each action type when different decision boundaries are used on the raters' aggregate preferences.}
\label{tab:count-alt-criteria}
\end{table}

\begin{table}[]
\begin{tabular}{|l|l|l|l|}
\hline
\multicolumn{4}{|c|}{Mean disagreement level}                                                                \\ \hline
Decision boundary               & Inform & Reduce & Remove \\ \hline
25\%           & 30.66  & 27.49  & 14.49  \\ \hline
33.33\%        & 26.89  & 23.31  & 12.28  \\ \hline
\textbf{50\% (default)} & 23.57  & 21.12  & 11.05  \\ \hline
66.66\%        & 25.34  & 22.78  & 11.55  \\ \hline
75\%           & 29.10  & 24.25  & 11.85  \\ \hline
\end{tabular}
\caption{Mean across articles, of percentage of raters who would disagree with the decision when different decision boundaries are used on the raters' aggregate preferences.}
\label{tab:disagreement-alt-criteria}
\end{table}

\rev{ Most of our analysis (particularly RQ\ref{rq-agreement}, RQ\ref{rq-ideology-amount}, and RQ\ref{rq-proxy}) used a majority-based decision boundary, i.e., the course of action against an article was decided based on whether a majority (>50\%) of raters wanted that action to be taken or not. In practice, however, there may be alternate decision criteria to determine the action(s) preferred by raters. For instance,  a ``one-yes technique'' in a team of three raters \cite{kohler2017supporting} will result in an action if any 1 out of 3 raters prefers that action to be taken. For a general case, that would mean setting the decision boundary at 33.33\% (i.e., 1/3). Similarly, one could experiment with other decision boundaries such as, 25\% (1 in 4 raters) or 66.66\% ( 2 in 3 raters).
\\
In this subsection, we provide additional results on how these decision boundaries impact -- 1) the number of articles that are recommended for each action type, 2) percentage of raters who disagree with the decisions, and 3) reducibility of the decision criteria in terms of misleading and harm judgments. First, Table \ref{tab:count-alt-criteria} shows how the numbers of articles recommended for each action type vary with five different decision boundaries -- 25\%, 33\% (inspired by the one-yes technique in a team of three), 50\% (our default majority-based criteria), 66\% (equivalent to a one-no technique in a team of three), and 75\%. As expected, less articles are recommended for each action as the decision boundary is increased. Interestingly, for our particular dataset, no article would be recommended for removal if the decision boundary was set at 75\%. 
\\
Similarly, Table \ref{tab:disagreement-alt-criteria} shows how the level of disagreement with the decisions varies based on various decision boundaries. 
The overall disagreement is minimized when the decision is based on a simple majority (i.e., > 50\%), and goes up on either side of this decision boundary. The disagreement levels increase more rapidly when the decision boundary is lowered and more articles are recommended for each action. This is expected as we found in Section \ref{sec:reducibility} that there is generally more consensus when the decision is to not take an action than otherwise.}

\begin{figure}
    \centering
    \begin{minipage}{0.33\textwidth}
        \centering
        \includegraphics[width=\textwidth]{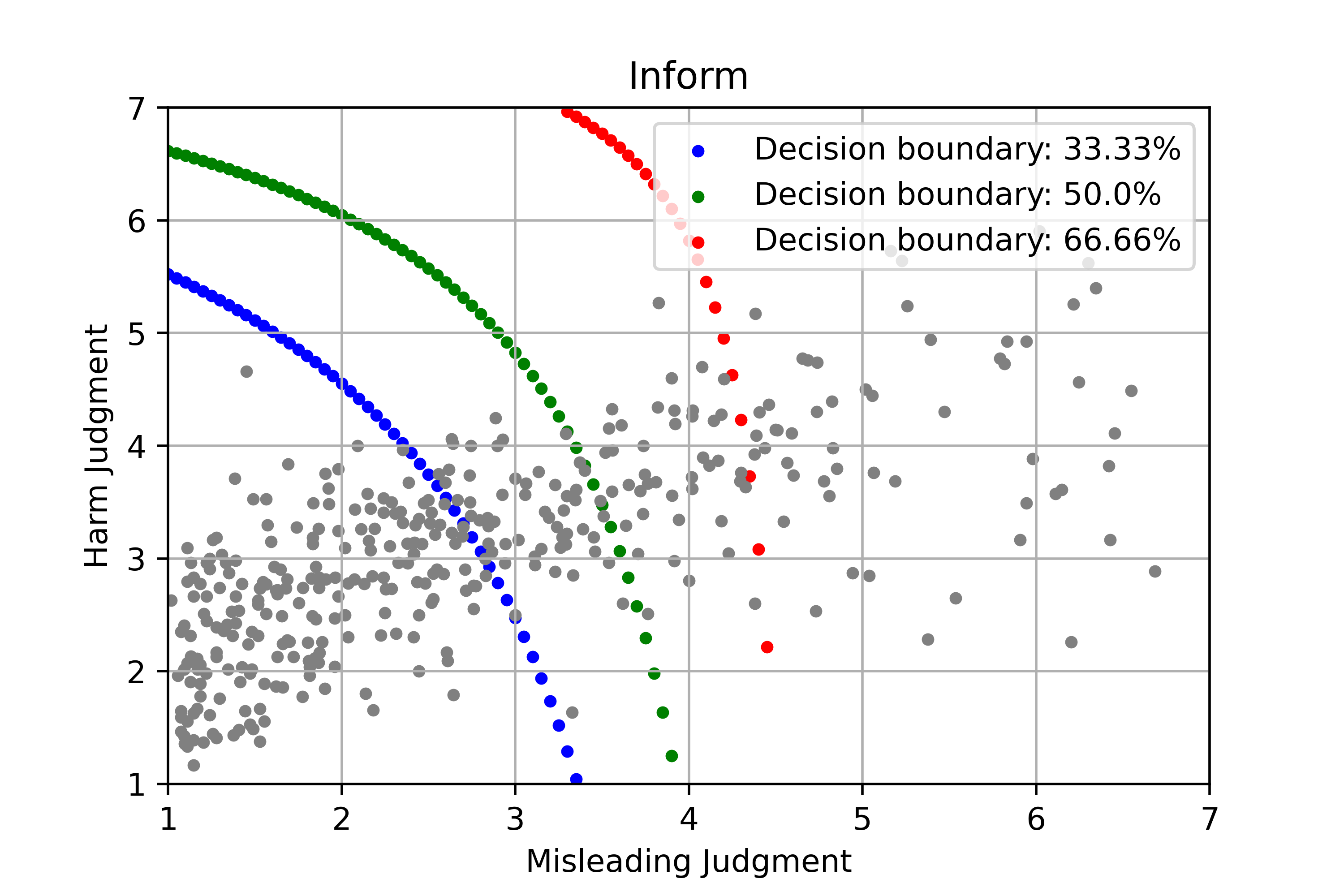} 
    \end{minipage}\hfill
    \begin{minipage}{0.33\textwidth}
        \centering
        \includegraphics[width=\textwidth]{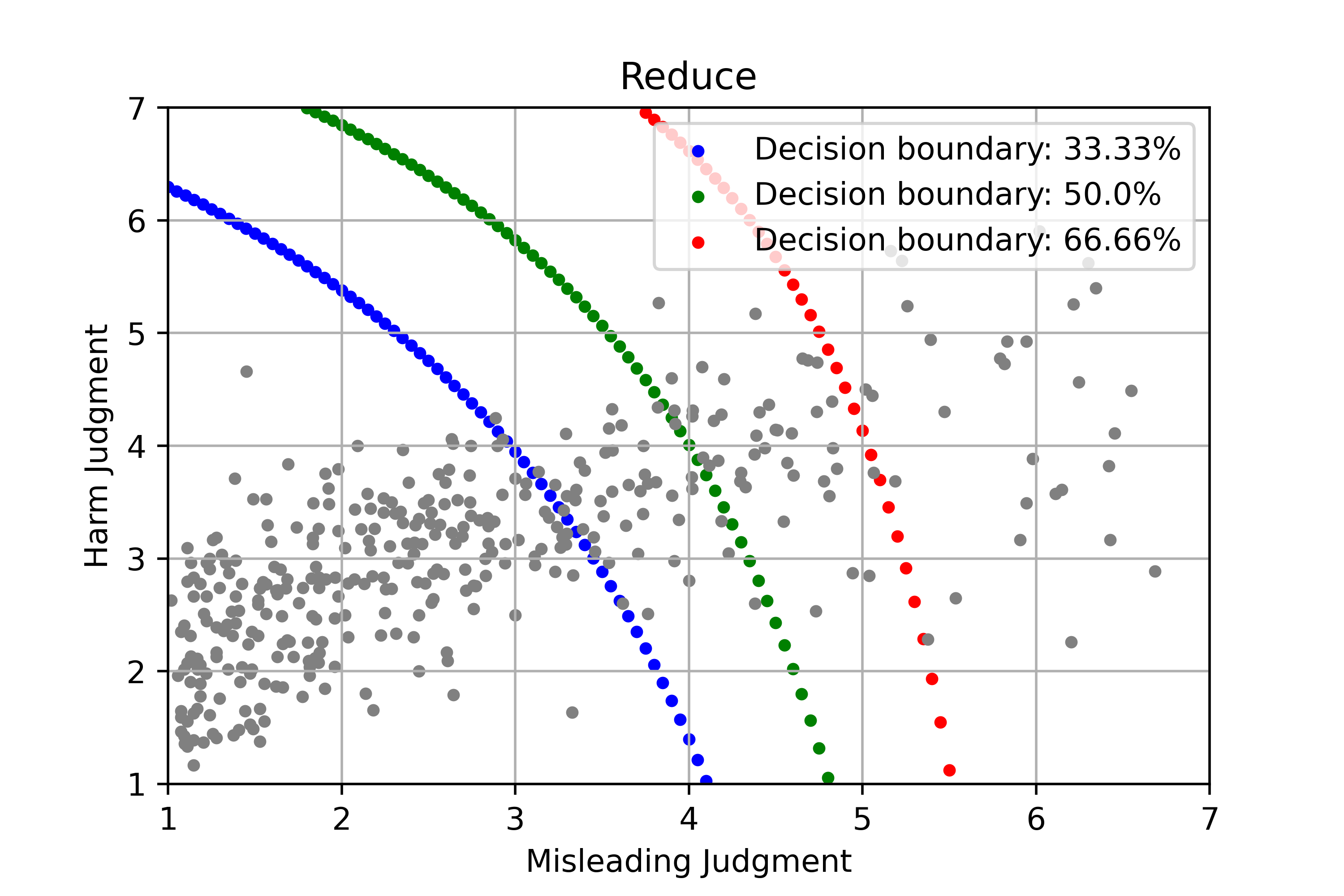} 
    \end{minipage}
    \begin{minipage}{0.33\textwidth}
        \centering
        \includegraphics[width=\textwidth]{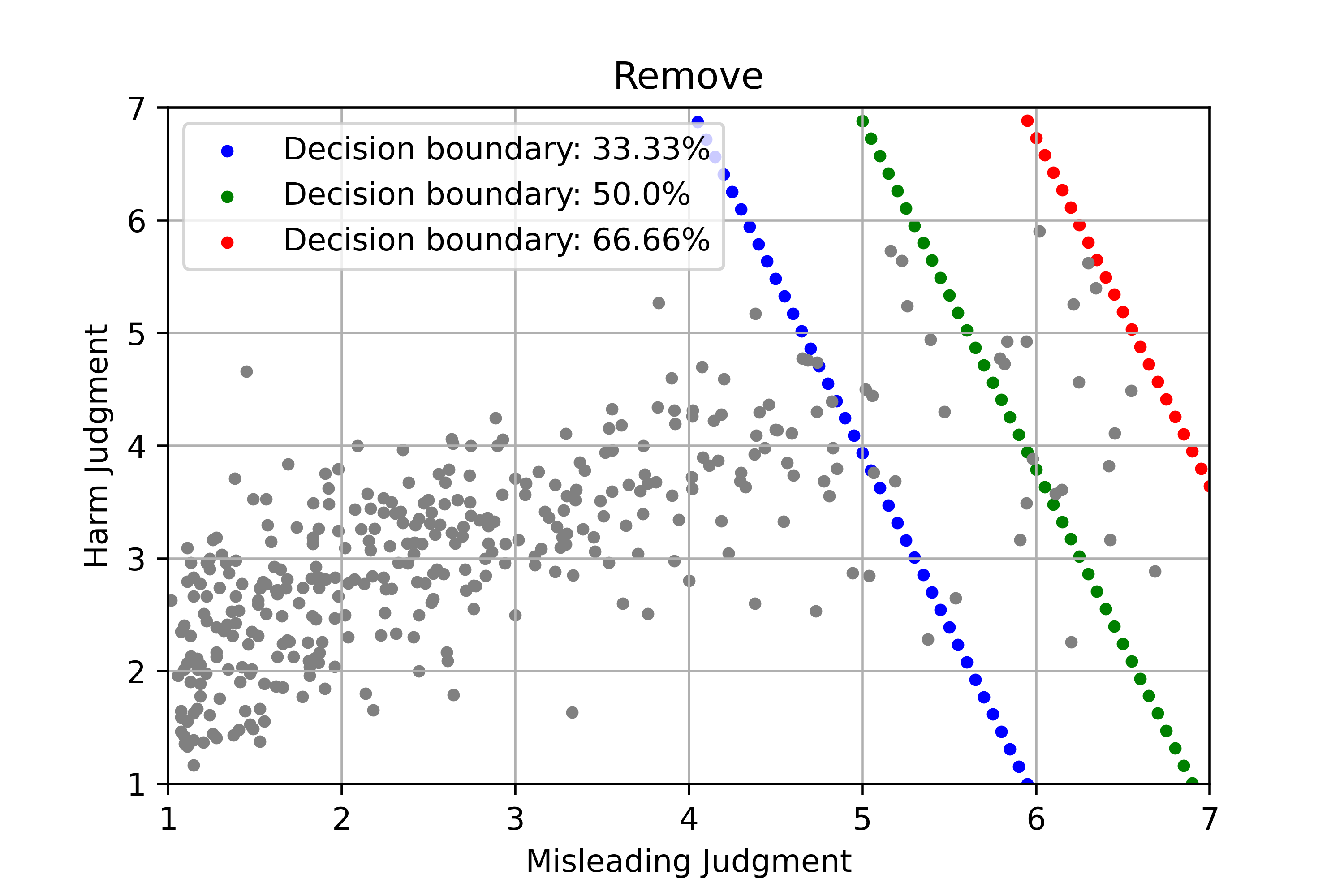} 
    \end{minipage}
    \caption{Visualizing the different decision boundaries in terms of misleading and harm judgments for inform, reduce, and remove actions (left to right)}
    \label{fig:reg-eq-alt-criteria}
\end{figure}

\rev{ Finally, Figure \ref{fig:reg-eq-alt-criteria} shows how these different decision boundaries map to the misleading and harm judgments. To improve the readbility of the figures, we only show 3 decision boundaries -- 33.33\%, 50\%, and 66.66\%. Observing individual data points in the figure, we see that articles with misleading and harm judgments of 3 and 3, respectively, would not be recommended for any action if the decision is based on simply majority, but would be recommended for inform if the decision boundary is set to 33.33\%. Furthermore, an article with misleading and harm judgments of 5 and 3, respectively, would no longer be recommended for reduce if the decision boundary was increased from 50\% to 66.66\% (it would still be recommended for inform in both cases).
\\
Overall, this analysis highlights how quantitative estimates of raters' action preferences vary with different decision criteria, while reinforcing our qualitative understanding of action preferences and their relationship to misinformation and harm judgments. 
}

\end{document}